\documentclass[twocolumn]{aastex63}
\usepackage[utf8]{inputenc}
\usepackage{amsmath}
\usepackage{tabularx}
\usepackage{amssymb}
\usepackage{lineno}
\usepackage{nameref}
\usepackage{array,multirow,makecell}
\usepackage{graphicx}
\usepackage{epstopdf}
\usepackage{color}
\usepackage{enumitem}
\usepackage{braket}
\usepackage[normalem]{ulem}
\usepackage{kantlipsum}
\submitjournal{ApJ}

\graphicspath{{./}{figures/}}
\begin{document}

\title{The detection of possible transient Quasi-Periodic Oscillations in the $\gamma$-ray light curve of PKS 0244-470 and 4C+38.41}%\footnote{Released on June, 10th, 2019}}

\author[0000-0002-9526-0870]{Avik Kumar Das}
\affiliation{Department of Physical Sciences, Indian Institute of Science Education and Research Mohali, \\
Knowledge City, Sector 81, SAS Nagar, Punjab 140306, India}

\author[0000-0002-1173-7310]{Raj Prince}
\affiliation{Center for Theoretical Physics, Polish Academy of Sciences, Al.Lotnikow 32/46, 02-668, Warsaw, Poland}

\author[0000-0002-9331-4388]{Alok C. Gupta}
\affiliation{Aryabhatta Research Institute of Observational Sciences (ARIES), Manora Peak, Nainital 263001, India}

\author[0000-0001-6890-2236]{Pankaj Kushwaha}
\altaffiliation{DST-INSPIRE Faculty Fellow}
\affiliation{Department of Physical Sciences, Indian Institute of Science Education and Research Mohali, \\
Knowledge City, Sector 81, SAS Nagar, Punjab 140306, India}
%\author[0000-0002-1188-7503]{Nayantara Gupta}
%\affiliation{Raman Research Institute, C. V. Raman Avenue, Sadashivnagar, Bangalore: 560080, India}

\email{avikdas@iisermohali.ac.in}
%We further tested the presence of QPOs over light curves extracted using different time-bin duration and found that the QPOs remain present.
%All the sub-structures except AP-3C show week scale periodicity with moderately local significance and five complete cycles. Similarly, the QP-1 phase of 4C+38.41 is divided into two sub-phases, defined as Q1 \& Q2 phases.
%We have also discussed the possible physical scenario responsible for QPOs of different time scales.
%All the detections are locally significant with at least four or more cycles.
\begin{abstract}
The continuous monitoring capability of Fermi-LAT has enabled the exploration of Quasi-Periodic Oscillations (QPOs) in the $\gamma$-ray light curve of blazar that has given a new perspective to probe these sources over a wide range of time scales. We report the presence of transient QPOs in the long-term $\gamma$-ray light curve of blazars PKS 0244-470 and 4C +38.41. We first identified different flux states using the Bayesian Block algorithm and then explored the possible transient QPOs in the segments of each flux phase where the flux level changes over fairly regular intervals. Combining this with the source's intrinsic variance, we identified two flux phases for PKS 0244-470: one activity (AP-1) and one quiescent phase (QP-1). For 4C+38.41, we similarly identified four activity (AP-1, AP-2, AP-3, AP-4) and two quiescent (QP-1, QP-2) phases. AP-1 phase of PKS 0244-470 shows QPO of $\sim$225 days persisting for 8 cycles ($\sim$ 4.1$\sigma$). In 4C+38.41, AP-1 and AP-2 phases show QPO-like behavior of $\sim$110 days and $\sim$ 60 days, respectively, persisting for 5 cycles. In AP-3, we identified three sub-phases, and all show a $\sim$ week scale possible recurrent rise with five complete cycles, while in QP-1, we could identify 2 sub-phases (Q1, and Q2). Q1 phase shows a period of $\sim$ 104 days with six complete cycles. Q2 phase also shows QPO but with only $\sim$3.7 cycles. We discuss the possible origin and argue that the current driven kink instability and curved jet model seem the most likely cause for shorter and longer QPOs.

%and found that all states of 4C+38.41 have QPO signature detected above 99.73\% local significance (except Flare-4, \& Flare-3C).
%Surprisingly, we also observe a significant QPO during the quiescent state with QPO time scale of $\sim$225 days.   
%\sout{The detection of QPOs in the same source during different flux states suggests the complex nature of the source and its surrounding.}

\end{abstract}


%%%%%%%%%%%%%%%%%%%%%%%%%%%%%%%%%%%%%%%%%%%%%%%%%%

%%%%%%%%%%%%%%%%% BODY OF PAPER %%%%%%%%%%%%%%%%%%
 %\sout {A relativistic jet is closely oriented along the observer's line of sight (\citealt{Urry1995})}.
\section{Introduction}
Blazars, a sub-class of active galactic nuclei (AGN), are among the most luminous and rapidly variable extragalactic sources in the Universe and have been observed in the entire accessible electromagnetic (EM) region (from radio to $\gamma$-ray/very high energy $\gamma$-ray). The entire emission is primarily from the relativistic jet and exhibits a characteristic broad double-humped spectral energy distribution \citep[SED;][]{1998MNRAS.299..433F} with one peak in-between near-infrared (NIR) to X-rays energies (low-energy hump) and the other at MeV-GeV energies (high-energy hump). The low-energy hump is widely accepted as the synchrotron emission from relativistic electrons within the jet while the high-energy hump origin is still debated and argued to be either due to emission from relativistic leptons and/or hadrons. In the leptonic case, it's due to inverse Compton scattering while in the hadronic scenario, plausible mechanisms are proton synchrotron and/or cascade initiated as a result of interaction with photons or particles. Traditionally, depending on the strength of optical emission lines, blazars have been categorized into two sub-classes: BL Lacertae (BL Lacs) objects and Flat spectrum radio quasars (FSRQs).\\
%\sout{Multi-wavelength SED (Spectral Energy Distribution) of blazars \textbf{are characterized by} broad-double hump structure \citep{1998MNRAS.299..433F}. \textbf{The one hump peaks at a low energy range (between IR to X-ray), which is usually described by the synchrotron emission of accelerated electrons in the jet. The other hump peaks at the high energy (HE) range, which covers the hard X-ray to Gev-Tev $\gamma$-ray part of the spectrum.} In the leptonic model, this HE hump can be explained by the inverse Compton (IC) emission of upscattered seed photons mostly originating in the synchrotron emission (SSC; \citealt{Sikora2009Sep}) and external photon fields: \textbf{BLR (Broad Line Region)}, CMB (Cosmic Microwave Background), etc. (\citealt{Dermer1992Mar}). Apart from this, a hadronic model can also explain this HE hump \textbf{by different hadronic processes such as proton-proton, proton-photon interactions, or proton synchrotron emission.} \textbf{BLLac objects are further divided into three categories based on the peak frequency ($\nu_{peak}$) of the low energy hump - Low energy peaked (LBL; $\nu_{peak} < 10^{14} Hz $), intermediate energy peaked (IBL; $10^{14} Hz < \nu_{peak} < 10^{15} Hz$) \& High energy peaked (HBL; $\nu_{peak} > 10^{15} Hz$)  BLLacs \citep[e.g.,][]{Tagliaferri2000Feb,2006A&A...445..441N,Massaro2008Feb,2010ApJ...716...30A,Barnacka2014Jul}.}} \\ 
\\ 
%{\bf !!! DIRECT INTRODUCTION OF X-RAY BINARIES WITHOUT INDICATING A CONNECTION WITH AGN/BLAZARS!!!} 
Blazars flux variation is primarily stochastic -- random and erratic \citep[e.g.][]{2014ApJ...786..143S}, with statistical behavior similar to the other accretion-powered sources over long-term \citep[e.g.][]{2016ApJ...822L..13K,2017ApJ...849..138K,2018RAA....18..141S}.  However, several light curves of blazars ($\gamma$-ray and other EM bands) show either transient or persistent (relative) quasi-periodic behavior. Since 2008, the detection or reporting of such QPOs in blazars, and in other sub-classes of AGN have been increased -- mainly due to much better data sampling as a result of coordinated MW follow-ups under the Fermi AGN Monitoring Program. Many strong QPOs have been reported across the complete EM bands in many sources on different time scales ranging from minutes to hours to days and to years {\citep[e.g.,][and references therein]{Gierlinski2008, Lachowicz2009, Gupta2009,2018A&A...616L...6G, Gupta2019, King2013,2014MNRAS.445L..16A,2015MNRAS.449..467A,2014JApA...35..307G,2018Galax...6....1G, Ackermann2015,2016ApJ...819L..19P, Zhou2018Nov,galaxies9020020,2021MNRAS.501.5997T}}.
%,Roy2022,Roy2022Mar
%,2021MNRAS.501...50S
%Transient Quasi-Periodic Oscillations (QPOs) have mainly been observed in galactic X-ray binaries and micro-quasars systems, where the accretion disk and corona mostly power the emission  \citep{2006ARA&A..44...49R}.
\\
%\textcolor{red}{QPO in blazar are also observed in radio and optical band by \citet{2001A&A...377..396R} and \citet{Roy2022}. They found the radio QPO of $\sim$ 5.7 years in the blazar AO 0235+164 and optical QPO of $\sim$ 8.13 years. An hour scale QPO in X-ray band was also observed in 3C120 by \citet{galaxies9020020}.}
%{\textcolor{blue}{\bf A few more sentences will be added here to explain QPOs on diverse periods except $\gamma-$ray bands.}} \\
%\sout{Until now, only in a handful of blazars, the QPOs are detected, but it has been proposed/claimed in many sources. 
%In the last decade,}
%\sout{wavebands}
%and challenging since the gamma-ray emission is produced in the highly collimated relativistic jet \citep{2017A&ARv..25....2P}
\\
The entire blazar emission, especially gamma rays are produced in the highly collimated relativistic jet, and as stated, the variability is primarily stochastic in nature. \citep{2010ApJ...722..520A, 2014ApJ...786..143S}. So, the QPOs reported in $\gamma$-ray are very interesting, indicating processes/mechanisms driving systematic changes than the usual flux variability, and thus, crucial in understanding not only jet physics but indirectly providing clues about acceleration mechanisms as well. The first QPO in the $\gamma$-ray was reported in blazar PG 1553+113 by \citet{Ackermann2015}, which was later confirmed by \citet{Tavani2018}. The QPO period was reported as 2.18 years, and three cycles were observed. Since then many have been QPOs in gamma-rays have been reported in other blazars, such as PKS 2155-304 by \citet{Sandrinelli2014}, where they detected a QPO of 1.73 years, which was later confirmed by \citet{Zhang2017}. The $\gamma$-ray QPOs of 3.35 yrs and 2.1 yrs were reported in blazar PKS 0426-380 and PKS 0301-243, respectively, in long period light curves (\citealt{Zhang2017b, Zhang2017a}). A systematic search for $\gamma$-ray QPOs on 3FGL source was done by \citet{Zhang2020}, and they detected a new source PKS 0601-70 with a possible QPO of 450 days. Apart from the strong detection of QPOs in many blazars, the study done by \citet{Covino2019} claims that the previously reported QPOs at different significance level in many blazars using the $\gamma$-ray light curve is basically not significant, and the red-noise highly dominates the power spectral density. The QPOs in the $\gamma$-ray light curve are also checked for a few BL Lacertae sources by \citet{Sandrinelli2018}, and they argue that if supermassive binary black hole system is the origin of QPOs in these sources, then there will be a tension with the upper limits on the gravitational wave background measured by the future pulsar timing array. Recently many more QPOs were reported in $\gamma$-ray in the bright AGN \citep[e.g.,][and references therein]{Ren2022Apr} and particularly in blazars \citep[e.g.,][and references therein]{2020A&A...642A.129S,2021MNRAS.501...50S,Gong2022}. \\
\\
%However, contrary to that, the blazar AO 0235+164 show a QPO of 2 years in its long-term optical light curve (\citealt{Roy2022}) standing with blazar OJ 287 where the $\sim$ 12 QPO is reported, and one of the possible explanation is proposed to be supermassive binary black hol system in these two sources. \\
%\\
PKS 0244-47 is an FSRQ type blazar with coordinate RA = 02h46m00.00s, DEC = -46d51m18.4s (J2000) located at redshift, z = 1.385 \citep{2010ApJ...715..429A}. It is identified as flaring $\gamma$-ray blazar in 2010 \citep{2010ATel.2440....1E} and was very active till 2013 (see the $\gamma$-ray light curve on \href{https://fermi.gsfc.nasa.gov/ssc/data/access/lat/LightCurveRepository/source.html?source_name=4FGL_J0245.9-4650}{Fermi repository}). Since 2013 the source is in a very low state, and few studies have been done. This is the first time we are presenting a QPO study on this source. \\
\\
The blazar 4C +38.41 ($\alpha_{2000.0} = \rm{16h \ 35m \ 15.4929s},  \delta_{2000.0} = +\rm{38}^{\circ} \ \rm{08}^{'} \ \rm{04.5}^{''}$)\footnote{\url{https://www.lsw.uni-heidelberg.de/projects/extragalactic/charts/1633+382.html}} is also an FSRQ, located at redshift, z = 1.814 \citep{Hewett2010Jul,Paris2018May}. %\citealt{Paris2018May}). 
\citet{Ciprini2009Jul} first reported high state activity of this source with gamma-ray flux (Fermi-LAT) of (1.38 $\pm$0.32) $\times$ 10$^{6}$ photons cm$^{-2}$ s$^{-1}$. It is also observed in different wavebands, from radio to hard X-rays, with the different ground and space-based telescopes, e.g., Effelsberg 100-m radio telescope, Guillermo Haro Observatory, Swift-XRT/UVOT (\citealt{Myserlis2012Oct}, \citealt{Raiteri2011Jul}, \citealt{Ghisellini2015Sep}). The blazar 4C +38.41 has been studied on a longer time scale considering the decade-long $\gamma-$ray data in \citet{Bhatta2020}, but they did not observe any QPO in this source. Recently, \citet{Ren2022Apr} studied 35 brightest sources and explored the QPO nature in the long-term $\gamma-$ray light curves ($\sim$12 yrs). They also did not notice any QPOs in 4C +38.41 in light curves extracted using 7 days and 30 days binning of data. \\
\\
Here, we present the variability study of the gamma-ray light curves of two blazars (PKS 0244-470 and 4C+38.41) and the detection of transient QPOs on various time scales. In section \S\ref{sec:data}, we discuss the gamma-ray light curve analysis procedure, and in section \S\ref{sec:3}, we discuss the methods used for QPO detection. In section \S\ref{sec:4}, we present the method to measure the significance of the detected QPOs, followed by results and discussion on plausible physical scenarios for transients QPOs in section \S\ref{sec:5} and section \S\ref{sec:discuss}, respectively. 

%\section{DATA ANALYSIS}

\section{Fermi-LAT data Analysis} \label{sec:data}

Fermi-LAT (Large Area Telescope) is a pair conversion and wide-FOV (Field of View) spaced-based gamma-ray telescope, working in an energy range between 20 MeV to $>$ 300 GeV \citep{Atwood2009May}. Detailed characteristic of the LAT is given in the Fermi-webpage\footnote{\url{https://fermi.gsfc.nasa.gov/ssc/data/analysis/documentation/Cicerone/Cicerone_Introduction/LAT_overview.html}}. It has an orbital period of $\sim$ 96 minutes and it alternately observes the northern and southern sky, thereby covering the entire sky in approximately three hours. \\
\\
We have analyzed the $\sim$ 13 years (December 2008 - December 2021) LAT data\footnote{\url{https://fermi.gsfc.nasa.gov/cgi-bin/ssc/LAT/LATDataQuery.cgi}} of two sources: PKS 0244-470 and 4C+38.41 (catalog name - 4FGL J0245.9-4650 and 4FGL J1635.2+3808 (\citealt{Ballet2020May}, \citealt{Abdollahi2020Mar})) between 100 MeV to 300 GeV using Fermi Science Tool software package \textbf{\it Fermitool} (version- 1.0.10). The standard selection criteria were chosen with \texttt{evtype=3} and the \texttt{evclass=128} to incorporate all types of photon-like events such as front, back, and front+back.
The light curve is extracted using a region of interest (ROI) of 10$^{\circ}$ around the source and a zenith angle cut of 90$^{\circ}$. The latter is chosen to reduce the contamination from the Earth's limbs. We have used the PASS8 data set re-processed with the instrument response function \texttt{P8R3\_SOURCE\_V6}. The source of interest is modeled using the Likelihood analysis implemented in the Fermi (pyLikelihood), and a \texttt{model.xml} file created from the Fermi Fourth source catalog (4FGL). The file also has sources beyond 10$^{\circ}$ ROI, but their parameters have been frozen to the catalog value during the analysis. The Power-law spectral model has been used for these two sources in the analysis. To account for the gamma-ray background, we have used the latest background models provided by the Fermi team, i.e., \texttt{iso\_P8R3\_SOURCE\_V6\_v06} for isotropic background and \texttt{gll\_iem\_v07} to account for the galactic diffuse emission. The strength of the $\gamma$-ray signal associated with the source position is characterized by the maximum likelihood analysis and by measuring the test statistics, TS = 2$\Delta$log(L), where L is the likelihood function for the models with and without a point source at the position of the source of interest. Further, to produce the light curve, we have fixed the parameters of all other sources within the ROI except our source of interest and generated the light curve for different bin sizes. We have also checked the outcome by freeing the spectral parameters of other variable sources within the ROI$\footnote{\url{https://fermipy.readthedocs.io/en/latest/install.html}}$ and found that it has no effect on the outcome. In further analysis, we have used only data points with high detection significance (TS $>$ 9). Figure-\ref{fig:1} shows the $\sim 13$-yr long gamma-ray photon flux history of the sources. All the reported $\gamma$-ray fluxes reported here are in units of 10$^{-6}$ ph cm$^{-2}$ s$^{-1}$.
%\sout{in the analysis}
\section{QPO DETECTION METHODS} \label{sec:3}
\subsection{Lomb-Scargle Periodogram}
Lomb-Scargle Periodogram (LSP) is one of the most well-known methods to search for periodicity in the unevenly spaced light curve. This method fits the sinusoidal wave to the time series data, a form of the least square method. The power of LSP is given by \citep{VanderPlas2018May} - 

\begin{equation}
  P = \frac{1}{2} \Big [\frac{{\sum_{i=1}^{N} x_i \sin{\Omega(t_i - \tau)}^2}} {{\sum_{i=1}^{N} \sin^2{\Omega(t_i - \tau)}}} + \frac{{\sum_{i=1}^{N} x_i \cos{\Omega(t_i - \tau)}^2}} {{\sum_{i=1}^{N} \cos^2{\Omega(t_i - \tau)}}} \Big ] 
\end{equation}
Here $\tau$ is - 

\begin{equation}
    \tau = \tan^{-1} \Big (\frac{\sum_{i=1}^{N} \sin{\Omega(t_i - \tau)}}{2 \Omega \sum_{i=1}^{N} \cos{\Omega(t_i - \tau)}} \Big )
\end{equation}

and $\Omega$ is the angular frequency ($\Omega = 2 \pi f$). 

In our work, we have chosen minimum ($f_{min}$) and maximum ($f_{max}$) value of the temporal frequency as $1/T$ and $1/(2\Delta T)$ respectively. $T$ is the total observation period for different sources. $\Delta T$ is the time duration used for extraction of one data point (e.g., 1, 2, and 10 days). We take total frequency interval as $N = n_0 T f_{max}$, where $n_{0} = 5$.\\
It is noted that several works have also used Generalized Lomb-Scargle Periodoram (GLSP)\footnote{\url{https://pyastronomy.readthedocs.io/en/latest/pyTimingDoc/pyPeriodDoc/gls.html}} for periodicity search, that takes into account the effect of measurement errors in the analysis. We have also analyzed the light curves with the GLSP method and found similar periods as the LSP method.

\subsection{Weighted Wavelet Z-transform}
Weighted Wavelet Z-transform (WWZ) is the most robust method to search transient periodicity in unevenly spaced time series data, which is relevant for most astronomical observations. This method decomposes the data into time and frequency domains (known as `WWZ Map'). By this procedure, we can detect periodicities in irregularly spaced light curve in a more sophisticated way than the well-known Discrete Wavelet Transform (DFT). DFT can give nontrivial statistical behavior even with regularly spaced data. For an extended discussion on this and the WWZ technique, we refer to \citet{Foster1996Oct}. 

The WWZ method is based on the weighted projection of the data vector onto the subspace, spanned by three trial functions,

\begin{equation} \label{eq:3}
%\phi_{i}(t) = 
    \begin{cases}
    
     \phi_{1}(t) = 1(t) \\
     \phi_{2}(t) = {\cos}(\omega(t-\tau)) \\
     \phi_{3}(t) = {\sin}(\omega(t-\tau)) 
     
    \end{cases}
\end{equation}

Where each trial function is an n-dimensional (length of the time series) vector:

\begin{equation}
    \boldsymbol{\phi_{i}(t)} = [\phi_{i}(t_{1}), \phi_{i}(t_{2}),...,\phi_{i}(t_{n})], \\ i = 1, 2, 3
\end{equation}

with statistical weight given by,

\begin{equation} \label{eq:5}
    \omega_{\alpha} = e^{-c\omega(t_{\alpha}-\tau)^2}, \\  \alpha = 1, 2, ..., n
\end{equation}

Here, $\omega$ and $\tau$ are the scale factor and time shift parameters, respectively. $c$ is known as the tuning parameter. In our study, we have chosen this value as 0.007. The projection coefficients ($y_i$) of the above three trial functions ($\phi_i$ in Equation-\ref{eq:3}) have been computed for which the model function ($ y(t) = \sum_{i=1}^{3} y_i\phi_i(t) $) best fits to the data vector. The best-fit coefficients are given by -

\begin{equation} \label{eq:6}
    y_i = \sum_{j} S_{ij}^{-1}\braket{\boldsymbol{\phi}_j|\mathbf{x}}
\end{equation}

where, $S_{ij}$ is the inverse of S-\textit{matrix}, defined by, $S_{ij} = \braket{\phi_i|\phi_j}$, and $x$ is time series data vector, $\mathbf{x} = [x(t_1), x(t_2) ..., x(t_n)]$.

We define the power or Universal Power Statistic (UPS) by the following formula \citep{Foster1996Jan}:

\begin{equation} \label{eq:7}
    P = \frac{N V_{y}}{(r-1)s^{2}}
\end{equation}
where, $V_{y}$ (= $\braket{y|y} - \braket{\bold{1}|y}^{2}$ = $\frac{\sum_{\alpha = 1}^{n} \omega_{\alpha} y^{2}(t_{\alpha})}{\sum_{\alpha = 1}^{n} \omega_{\alpha}} - [\frac{\sum_{\alpha = 1}^{n} \omega_{\alpha} y(t_\alpha)}{\sum_{\alpha = 1}^{n} \omega_{\alpha}}]^{2}$) is the weighted variations of the model function and can be calculated using equation-\ref{eq:6} (i.e., $ y(t) = \sum_{i=1}^{3} y_i\phi_i(t) $) and equation-\ref{eq:5}. $r$ and $N$ are the numbers of the trial functions used for the projection and data points in the given time series, respectively. $s^{2}$ (= $\frac{N V_{x}}{N-1}$) is the estimated variance of the data with $V_x = \braket{x|x} - \braket{\bold{1}|x}^{2} = \frac{\sum_{\alpha = 1}^{n} \omega_{\alpha} x^{2}(t_{\alpha})}{\sum_{\alpha = 1}^{n} \omega_{\alpha}} - [\frac{\sum_{\alpha = 1}^{n} \omega_{\alpha} x(t_\alpha)}{\sum_{\alpha = 1}^{n} \omega_{\alpha}}]^{2}$ , weighted variation of the data vector. When we treat wavelet transform as weighted projection, we use effective number of data points (i.e., $N_{eff} = \frac{(\sum_{\alpha = 1}^{n} \omega_{\alpha})^{2}}{\sum_{\alpha = 1}^{n} \omega_{\alpha}^{2}}$) instead of $N$ (ref section-4.3 of \citealt{Foster1996Jan}) to compute the power. So, using r=3 and $N = N_{eff}$ in equation-\ref{eq:7}, we get the Weighted Wavelet Transform (WWT): 

\begin{equation}
    WWT = \frac{(N_{eff} - 1)V_y}{2V_x}
\end{equation}

%Where $V_x$ (= $\braket{\bold{x}|\bold{x}} - \braket{\bold{1}|\bold{x}}^{2}$) is the weighted variations of the data. 
However, this quantity is highly sensitive to the $N_{eff}$ and causes a false peak at low frequency. So instead of WWT, we use Z-statistic for projection \citep{Foster1996Jan}, named as Weighted Wavelet Z-transform (WWZ) - 

\begin{equation}
    WWZ = \frac{(N_{eff} - 3)V_y}{2(V_x - V_y)}
\end{equation}
which follows F-distribution with degrees of freedom $N_{eff}-3$ and 2. 
%WWZ maps of 4C+38.41 for different phases are illustrated in Figure-4, Figure-7 \& Figure-10.
\begin{figure*}[t!]
\centering
\includegraphics[width=\textwidth,height=8cm]{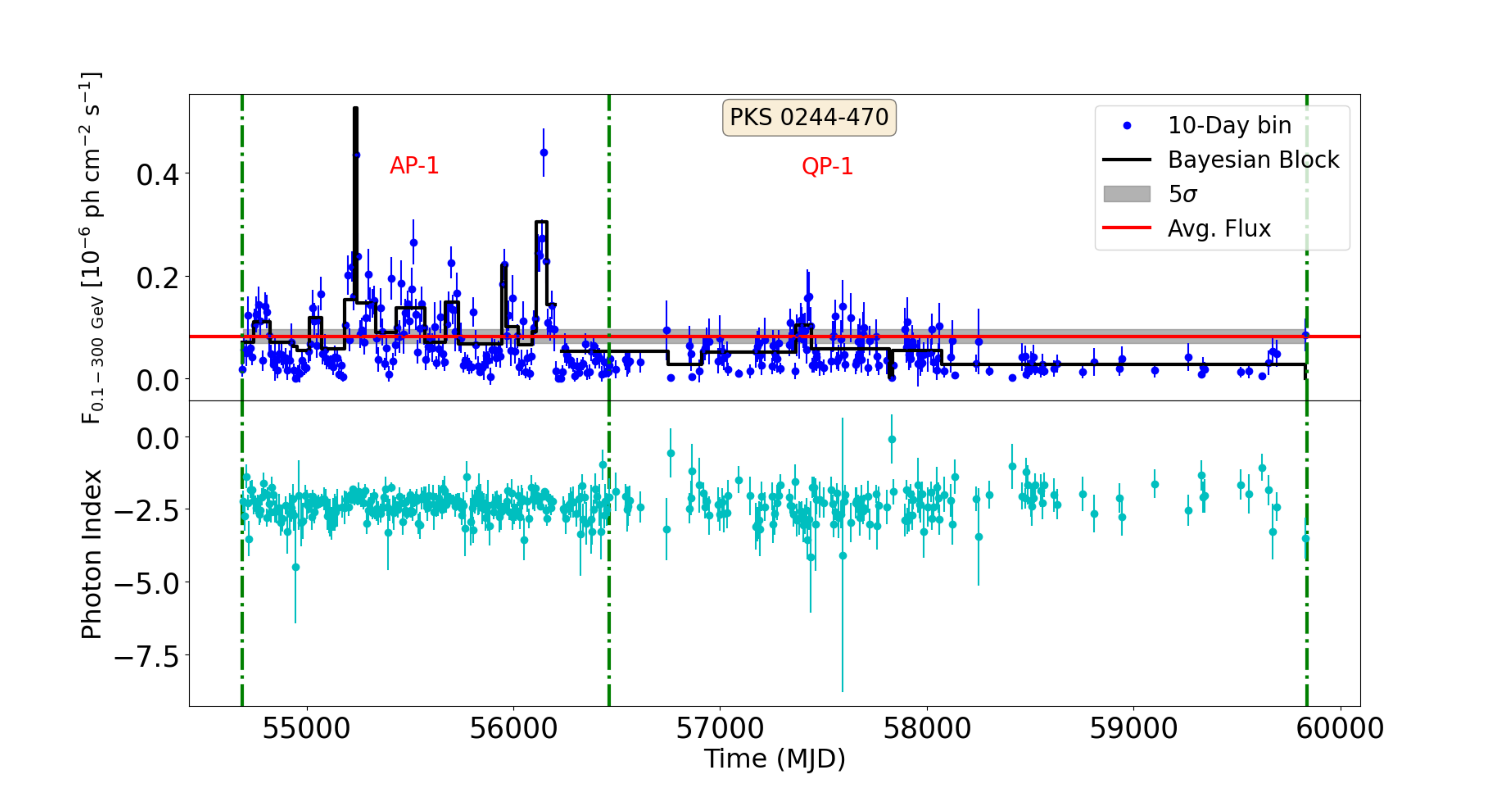}
\includegraphics[width=\textwidth,height=8cm]{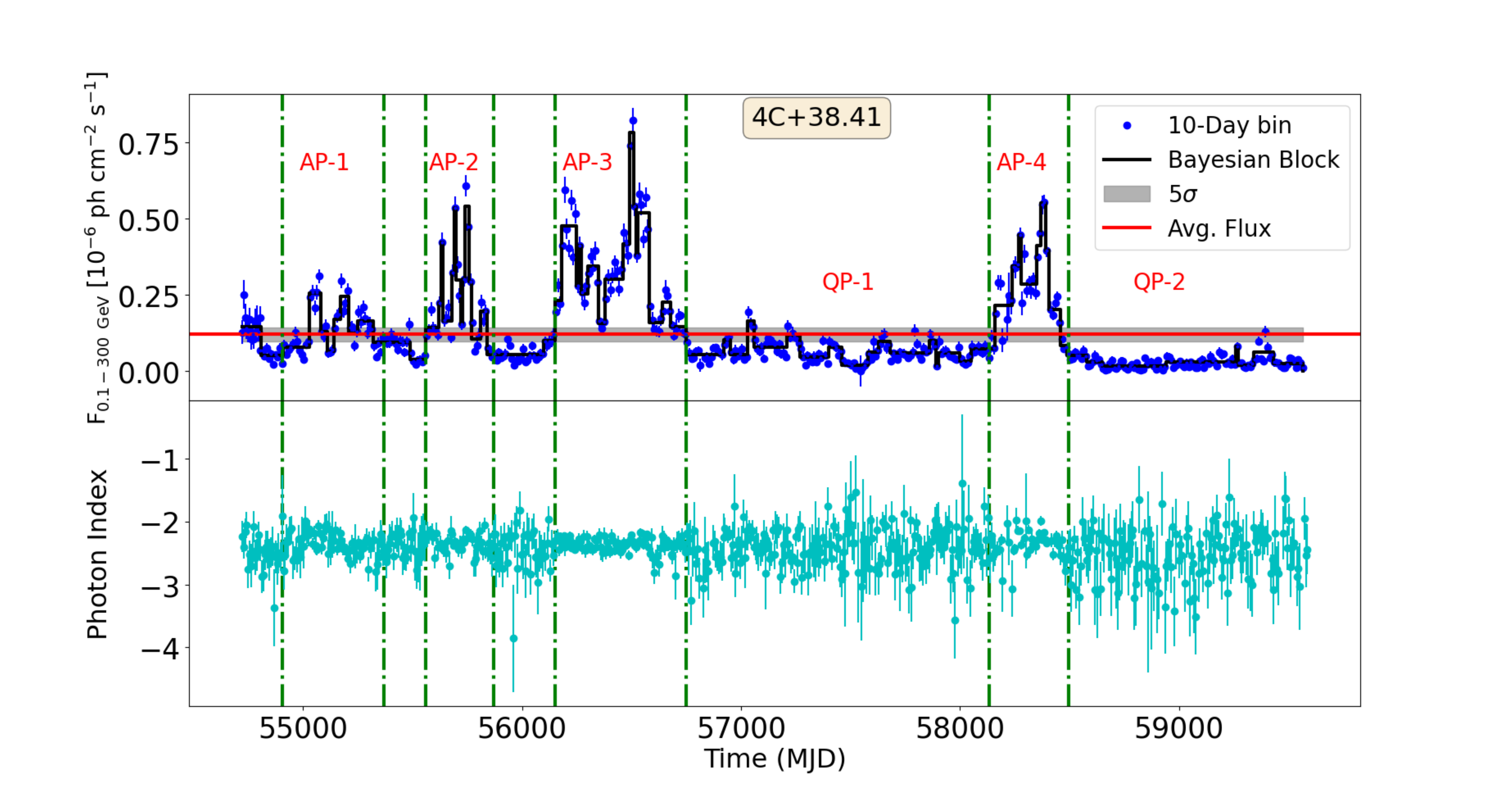}
\caption{10 days binned (time duration of $\sim$ 13 years) Fermi-LAT light curve of PKS 0244-470 (upper panel) and 4C+38.41 (lower panel). Different phases of activity are shown by the dash-dot green vertical line. The photon index values are also shown simultaneously below each of the light curves by cyan-colored points. Data points with lower detection significance (TS$<$9) are not shown in the plots.}
%95\% confidence contour is also shown in magenta color. 
%\textbf{(ref Figures \ref{fig:3}, \ref{fig:5}, \ref{fig:6})}
\label{fig:1}
\end{figure*}
\section{Significance estimation} \label{sec:4}
Though LSP and WWZ methods show recognizable peaks in the power plots, the statistical properties of the blazar light curve exhibit a red-noise process (power-law type). Due to the presence of this noise (it can be shown from the observed periodogram and auto-correlation function of the time series), the light curve can show periodic behavior of few cycles in the low-frequency regime (\citealt{Press1978}, \citealt{Vaughan2005Feb}). Therefore red-noise behavior should be appropriately considered when estimating the significance of the peak observed in the periodogram. 

To estimate the significance of the periodicity detection, we have used the Power Spectrum Response Method (PSRESP; \citealt{Uttley2002May}), which has been used extensively to model the periodogram (e.g. \citet{Chatterjee2008Dec}; \citet{Edelson2014Nov}; \citet{Bhatta2016Oct}, \citet{Benkhali2020Feb}). 

We first modeled the observed periodogram with a power law model -
\begin{equation}\label{eq:10}
    P(\nu) \propto {\nu}^{-\beta} + C
\end{equation}
%P(\nu) = A(\frac{\nu}{\nu_0})^{-\beta} + C
where $\beta$ is the spectral index of the model. C represents the Poissonian noise level, which is given by \citep{Bhatta2019Aug} - 

\begin{equation}
    C = \frac{2T<F_{err}^2>}{N^2 \mu^2}
\end{equation}

Here N is the number of data points during the time span of observation T. $\mu$, and $<F_{err}^2>$ describe the average flux and mean square of the flux errors, respectively. We are interested in finding the best-fitted spectral index ($\beta$) of the given PSD model. To find this, we simulate 1000 light curves by \citet{Timmer1995Aug} algorithm for each value of $\beta$ between 0.1 to 2.0 with step size 0.1. After that, we calculate the observed and re-sampled simulated periodograms to compute the following $\chi^2$-like quantities:

\begin{equation}\label{eq:12}
    \chi_{obs}^2 = \sum_{\nu = \nu_{min}}^{\nu_{max}} \frac{[<P_{sim}(\nu)> - P_{obs}(\nu)]^2}{<\Delta P(\nu)_{sim}>^2}
\end{equation}

\begin{equation}\label{eq:13}
    \chi_{sim,i}^2 = \sum_{\nu = \nu_{min}}^{\nu_{max}} \frac{[<P_{sim}(\nu)> - P_{i}(\nu)]^2}{<\Delta P(\nu)_{sim}>^2} 
    , (i = 1,2,...,1000)
\end{equation}

\begin{figure*}[t!]
\centering
\includegraphics[width=\textwidth,height=8cm]{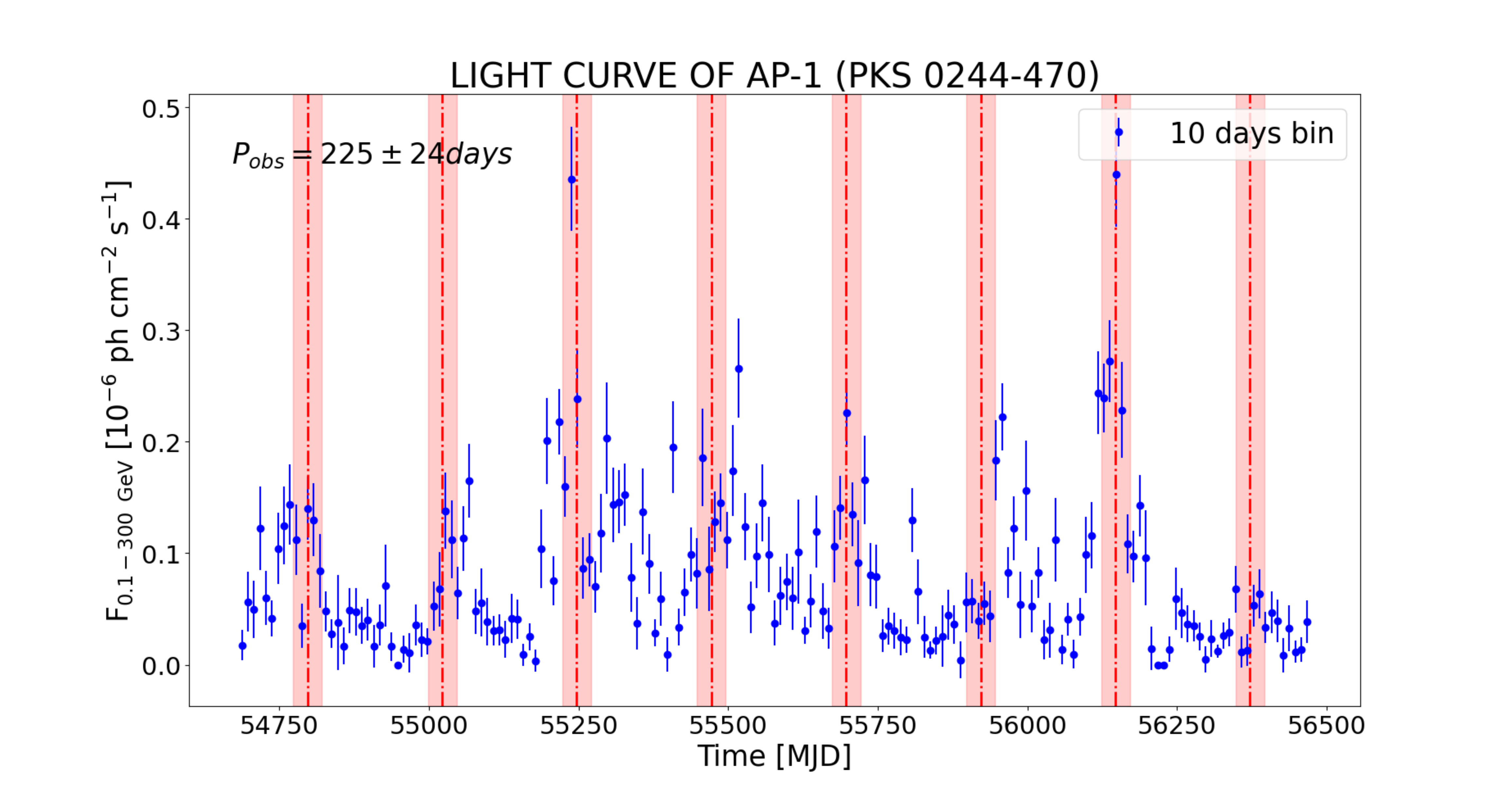}
\caption{Fermi-LAT light curve of AP-1 phase (MJD 54682 - 56475) of the blazar PKS 0244-470 in 10-day time bins. The vertical dash-dot red lines and red-shaded regions indicate the peak positions of the periodic oscillation and uncertainty on the peaks, respectively.}
%95\% confidence contour is also shown in magenta color. 
\label{fig:2}
\end{figure*}

where, $\nu_{min}$ and $\nu_{max}$ are the minimum ($\frac{1}{T}$) and maximum ($\frac{N}{2T}$) temporal frequencies of the periodograms. $<P_{sim}(\nu)>$ and $<\Delta P(\nu)_{sim}>$ are the mean and standard deviation of the simulated periodograms at a given frequency. Powers in the periodogram are not Gaussian variables, hence the above quantities (Equation-\ref{eq:12}, and Equation-\ref{eq:13}) are not the same as usual $\chi^{2}$ distribution \citep{Uttley2002May}. To quantify the goodness of fit, we have computed the success fraction for each spectral index ($\beta$)\footnote{\url{https://github.com/avikdas4/avikdas4}}. This is defined by the ratio of number of $\frac{\chi_{sim, i}^2}{\chi_{obs}^2} > 1$ to the total number of $\chi^{2}_{i}$ value for a given spectral index. The $\beta$-value for which the success fraction is maximum best represents the observed periodogram. We have fitted the results of success fraction vs. $\beta$ with a Gaussian function to estimate the best-fitted spectral index.

After modeling the source periodogram using equation-\ref{eq:10}, we simulated 10000 light curves for the best-fitted value of $\beta$ (corresponds to the peak of the Gaussian function) and performed the LSP method for each simulated light curve to estimate the significance level of the detection. We have also computed the significance level of WWZ peaks by a similar method. 

%Results of LSP and WWZ for different periods of activity (flare-1 \& flare-2) have been shown in Table-\ref{tab:1}.

\section{Results}\label{sec:5}

\subsection{Identifying different time segments}
We have used Bayesian Block (BB) representation \citep{Scargle2013Feb} to identify different flux states of activity. An Activity Phase (AP) is a sequence of consecutive recurring enhancement states, in which the BB flux level (represented by the black solid line in Figure-\ref{fig:1}) of every enhancement state crosses above the 5$\sigma$ standard deviation of the average flux value. In a few cases, the first or last cycle of the AP phase (e.g., for PKS 0244-470: the last cycle of AP-1 phase) has a low BB flux level ($<\mu+$5$\sigma$; $\mu$ = Average flux of the whole light curve) but follows a similar quasi-periodic variability pattern as the identified phase. In these cases, that time duration/cycle has been considered as part of the phase for the periodicity analysis. \\
%The flaring state is defined by block in which the flux level, crosses above the 5$\sigma$ standard deviation of the average flux value.
%, and then combining with mean flux and error, have identified different activity phases in both the sources.
%\red{(NOT True! Phase is consecutive.. flaring state is a duration when flux crosses 5-sig)}
%\sout{Activity Phases (APs) are further defined by the combinations of these consecutive flaring states}
\\
Through the above-mentioned procedure, we identified a $\sim$ 4-year prolong activity phase AP-1 (ref upper panel of Figure-\ref{fig:1}) followed by a similar prolong quiescent phase (QP-1) in PKS 0244-470 with apparently quite a regular flux changes during the AP phase. Similarly, we identified four high activity or AP phases (AP-1, AP-2, AP-3, and AP-4; ref lower panel of Figure-\ref{fig:1}) and two quiescent phases (QP-1 and QP-2) in 4C+38.41 with an indication of regular flux changes in both the phases. 
\subsection{QPOs search results}
%To further explore possible QPOs,
We exploited two of the most extensively used time series methodologies: LSP and WWZ based on two different underlying governing principles. We applied the significance estimation method, `PSRESP' (ref \S\ref{sec:4}) on each phase of light curves with different binning (1 day, 2 days, 5 days, and 10 days) criteria. Depending on the uncertainty in photon counts and the detailed structure of the light curve, we use different binning criteria for different phases to explore further. For example, the AP-1 phase of 4C+38.41 shows significant QPO ($>$ 99.99\% in both LSP and WWZ methods) at $\sim$ 110 days in light curve extracted using one-day binning, but due to a large error and many low TS data points (many data points have $F_{i}$ $\le$ $e_{i}$ or TS $\le$ 9 or both), we instead used two-day binning for the AP-1 phase. For similar reasons, we have shown the result of the AP-1 phase of PKS 0244-470 and QP-1 phase (Later, we again divide this phase into two different sub-phases: Q1 and Q2 phases) of 4C+38.41 in 10 days time bin. \\ 
%In order to estimate the significance of the QPO peak, we first modeled the observed PSD with a power-law function (ref second column of Table-\ref{tab:1}) and simulated 10000 light curves by \citet{Timmer1995Aug} algorithm for the best fitted PSD. \\
%\red{The significance of the peaks, if any, were then estimated by modeling the}
\begin{figure*}
\centering
\includegraphics[width=\textwidth,height=8cm]{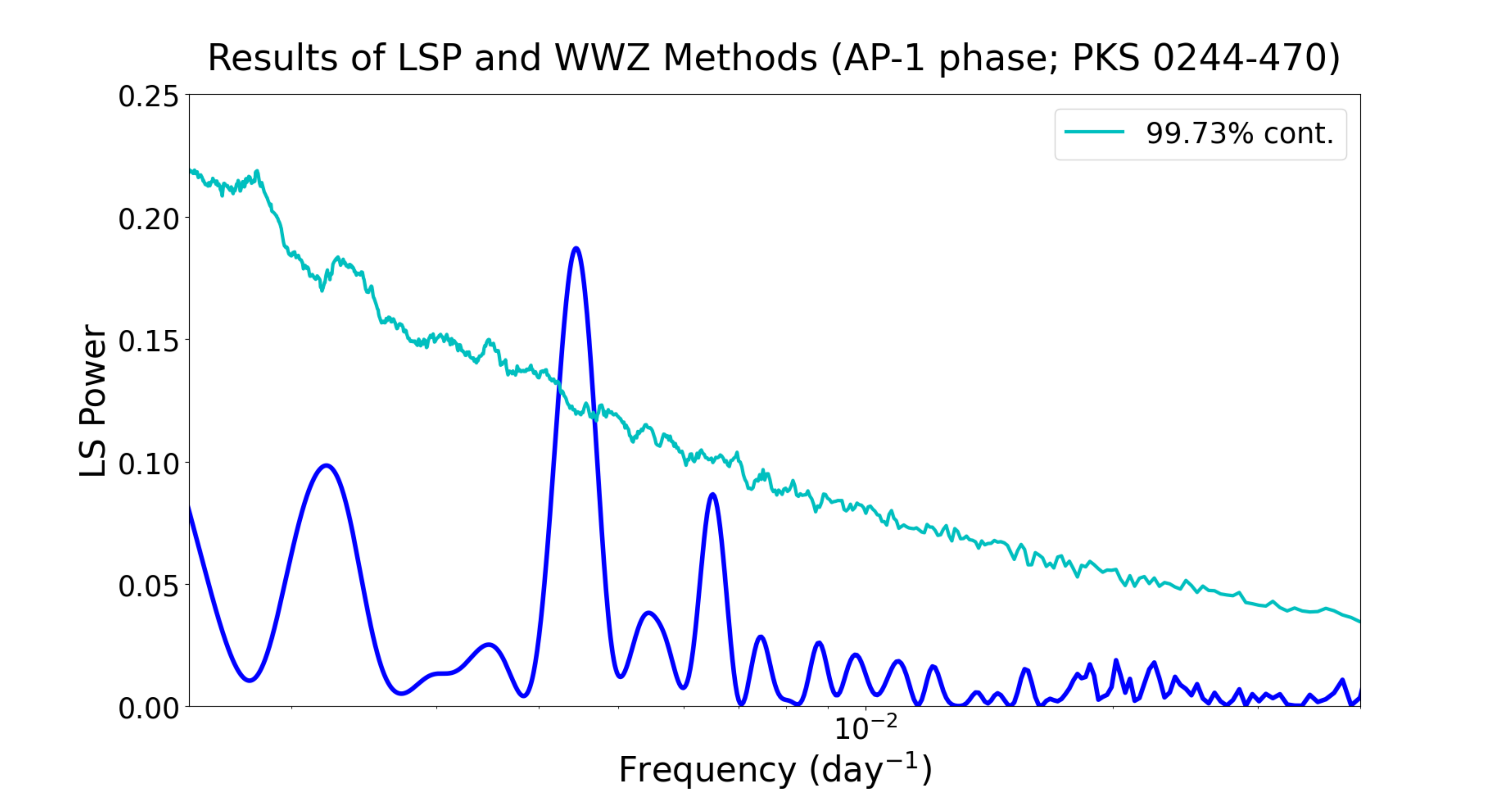}
\includegraphics[width=\textwidth,height=8cm]{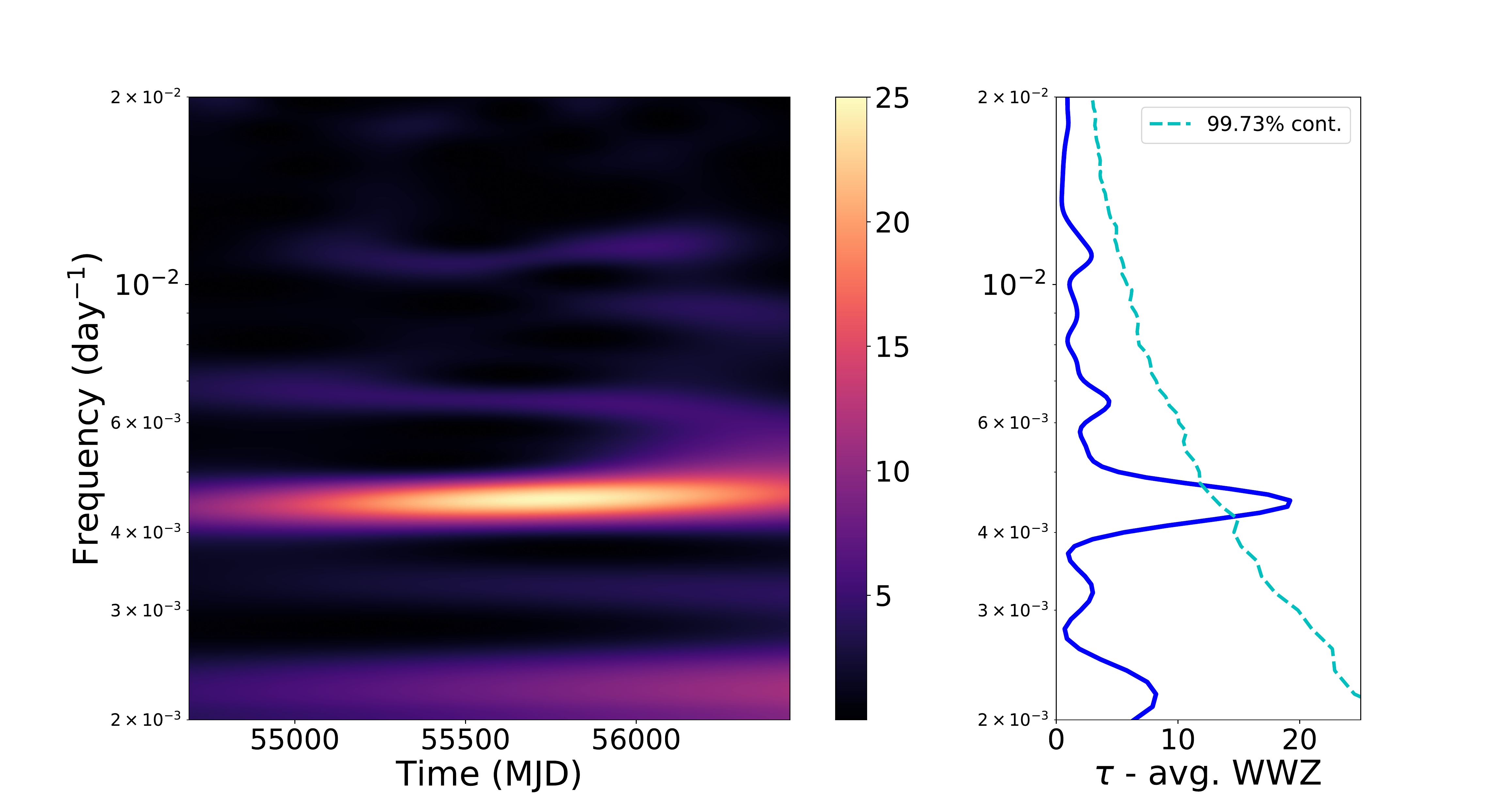}
\caption{Results of LSP (upper panel) and WWZ (lower panel) methods for AP-1 phase (PKS 0244-470). \textbf{ Lower-panel:} The left side image shows the WWZ map. The right side image shows the time-averaged WWZ power as a function of frequency (day$^{-1}$). The cyan color curve represents the 99.73\% local significance level.}
%95\% confidence contour is also shown in magenta color.
\label{fig:3}
\end{figure*}
%\sout{namely, AP-1 \& one Quiescent state for PKS 0244-470 and AP-1, AP-2, AP-3, AP-4, and two Quiescent-states (Q1 \& Q2) for 4C+38.41. Two different methods, Lomb-Scargle Periodogram (LSP) and Weighted Wavelet Z-transform (WWZ), have been used to detect possible transient QPOs in each activity states.
%\\

%AGN light curves show a red noise spectrum in general and, and hence extracting any periodic or quasi-periodic behavior from them is very challenging \citep{Vaughan2005Feb}. A robust estimation of significance is necessary to avoid claim of a few recurrence as QPOs which is typical of time series governed by red noise. This becomes more critical in transient QPOs and systems where constituents physical entities emit prominently in specific energy band - enabling such features in one band but not in other e.g., like AGNs, accretion-powered sources etc. Indirectly it allows probe of system's constituents shadowed otherwise by the jet emission and the exploration of underlying physical processes.\\ 

\begin{figure*}
\centering
\includegraphics[width=1.08\textwidth,height=0.7\textheight]{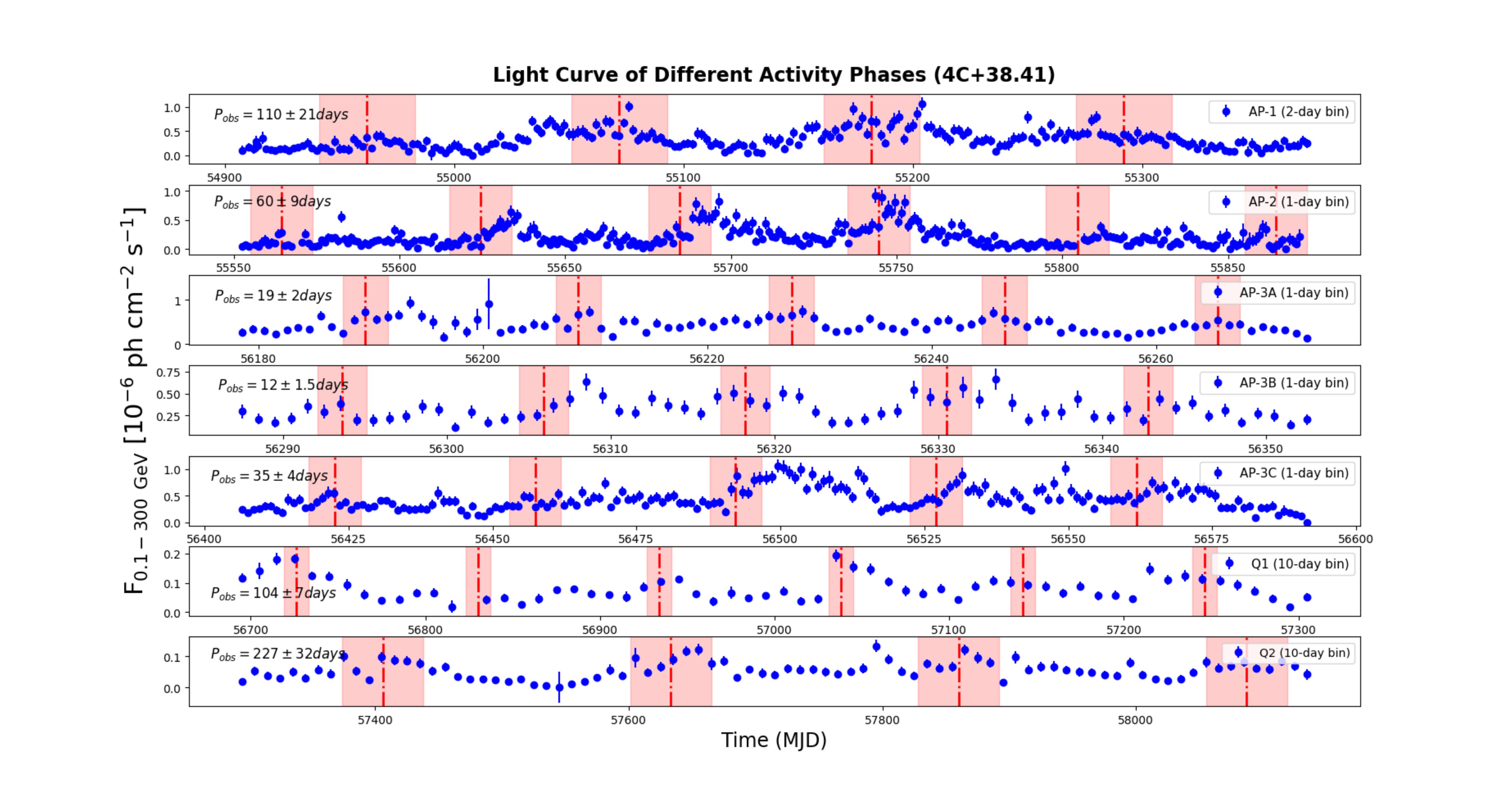} %[width=20.5cm,height=14.0cm]
\caption{Fermi-LAT light curve of different activity phases (AP-1, AP-2, AP-3A, AP-3B, AP-3C, Q1 and Q2 phases) of the blazar 4C+38.41. The upper right box of each panel shows the time bin values used for the light curve extraction and further analysis. The vertical dash-dot lines and red-shaded regions indicate the peak positions of the periodic oscillation and uncertainty on the peaks for each phase, respectively.}
%95\% confidence contour is also shown in magenta color. 
\label{fig:4}
\end{figure*}

\begin{figure}
\centering
\includegraphics[width=0.5\textwidth,height=4cm]{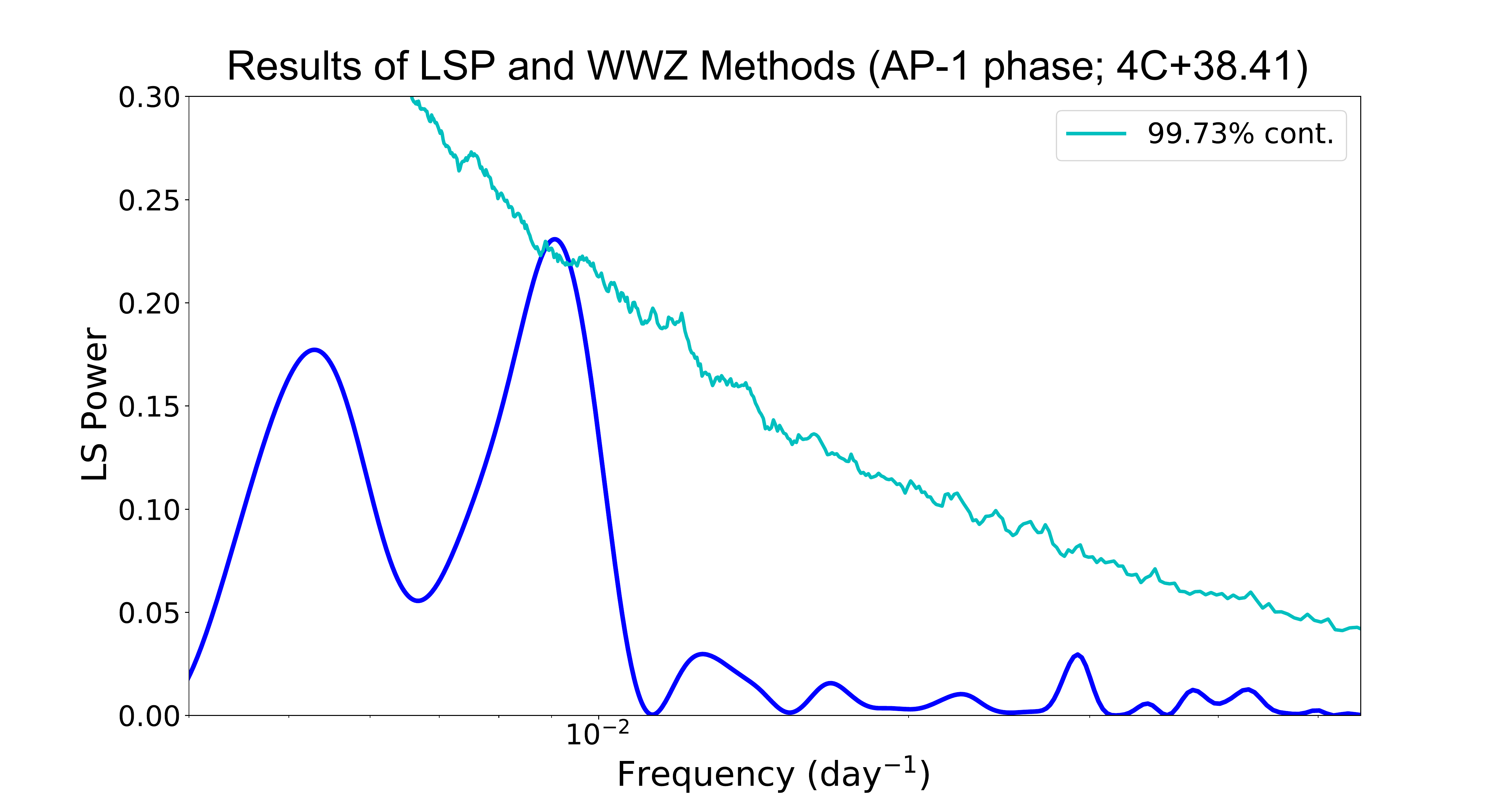}
%\caption{Lomb-Scargle Periodogram (LSP) of Flare-1 of 4C+38.41. The right side image shows the time averaged LSP power as function of frequency (day$^{-1}$).}
\includegraphics[width=0.5\textwidth,height=4cm]{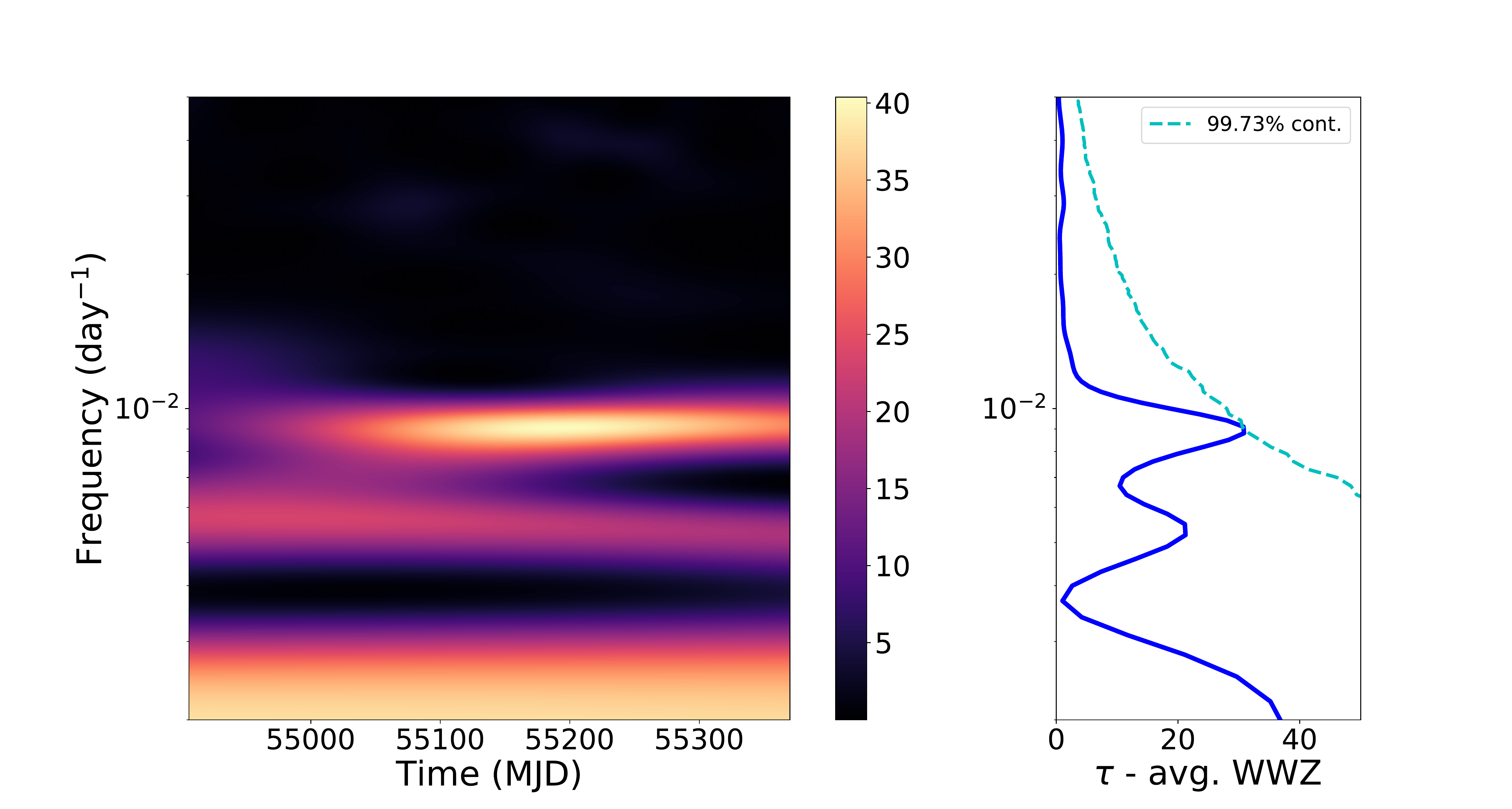}
\caption{Results of LSP (upper panel) and WWZ (lower panel) methods for AP-1 phase (4C+38.41). The left side image (lower panel) shows the WWZ map. \textbf{Lower-panel:} The right side image shows the time-averaged WWZ power as a function of frequency (day$^{-1}$). The cyan color curve represents the 99.73\% local significance.}
%95\% confidence contour is also shown in magenta color. 
\label{fig:5}
\end{figure}

%\sout{Here we discussed our results for both the sources (PKS 0244-470 \& 4C+38.41) and found that transients QPOs are above the 99.73\% local significance level for all the phases (except AP-3C of 4C+38.41) and have been detected for four or more cycles.}\\
%%\\
%The long term light curve of PKS 0244-470 is divided into two phases: AP-1 (MJD 54685 - 56475) \& Quiescent or QP-1 phase (56475 - 59565), which is shown in Figure-\ref{fig:1}.
Our study of high activity phase AP-1 (MJD 54685 - 56475; ref Figure-\ref{fig:2}) of PKS 0244-470, results in a possible QPO of 225$\pm$24 days with a significance of 99.996\% and 99.986\% in LSP and WWZ methods (ref Table-\ref{tab:1}) persisting for 8 cycles. The uncertainty on the periods reported here is estimated using half-width at half-maximum (HWHM) of the LSP result following \citet{VanderPlas2018May}. We caution that such a measure is not meant for broad QPOs. This phase is very prominent till the 7th cycle and weakens in the 8th cycle followed by its disappearance afterward. This is also the most significant QPO signal reported in this work. The LSP and WWZ plots are shown in Figure-\ref{fig:3}. The 99.73\% local significance contours are also shown by dashed cyan color in the LSP and time-averaged ($\tau$ - avg.) WWZ plots. \\
\\
Similarly, we have also searched for transient QPO-like variations in all the phases (AP-1, AP-2, AP-3, AP-4, QP-1, and QP-2) of 4C+38.41 as marked by dashed-dot green lines in the lower panel of Figure-\ref{fig:1}. The light curves of each phase with exact peak positions of oscillation (by dashed-dot red lines) and their corresponding uncertainties (by red shaded regions) on the periods have been shown in Figure-\ref{fig:4}. \\ 
\\
In 4C+38.41, AP-1 has a total time span of 463 days (MJD 54907 - 55370). It shows a periodicity of 110$\pm$21 days (2-day bin) with four complete cycles with the significance of 99.82\% and 99.77\% in LSP and WWZ methods respectively (ref Figure-\ref{fig:5}).
AP-2, on the other hand, has a duration of 319 days (MJD 55552 - 55871) and shows a QPO-like variation of 60$\pm$9 days with 5 complete cycles. The detected significance of the peak is 99.90\% in LSP and 99.85\% in the WWZ method.
%\red{!! ALREADY TALKING ABOUT IT SO WHY THIS? !!}.
\begin{table*}[!ht]
\caption{Results of LSP and WWZ method for different activity phases. Uncertainty on the PSD-slopes result from the HWHM (Half width at half maximum) of the Gaussian fit.} %title of the table
\centering
\begin{tabular}{ccccc rrrr}   % creating eight columns
\\
% & & Flare-A\\
\hline\hline                        
Activity-Phases & PSD-Slope & Detection Methods & Observed Period & No. of cycles & Detection Significance \\ [0.8ex] %&  Detection Significance\\ [0.8ex] % inserts table 
& [$\beta$] &  & [days] & & (local) \\ %& (global)\\
\hline\hline
& & & PKS 0244-470 & \\ %& \\
\hline\hline
AP-1 &  0.67$\pm$0.21 & LSP & $\sim$225 & 8.0 & 99.996\% \\ %& 98.164\% \\
& & WWZ & $\sim$222 & 8.0 & 99.986\% \\ %& 99.941\% \\
\hline\hline % inserts single-line
& & & 4C+38.41 & & \\
\hline\hline
AP-1 & 0.97$\pm$0.29 &  LSP & $\sim$110& 4.2 & 99.82\% \\ %& 91.46\% \\% Entering row content
& & WWZ & $\sim$111 & 4.2 & 99.77\% \\ %& 97.85\% \\
\hline
AP-2 &  0.83$\pm$0.12 & LSP & $\sim$60 & 5.4 & 99.90\% \\ %& 95.10\% \\
& & WWZ & $\sim$60& 5.3 & 99.85\% \\ %& 99.67\%\\
\hline
AP-3A &  0.60$\pm$0.29 & LSP & $\sim$19 & 5.1 & 99.98\% \\ %& 98.75\% \\
& & WWZ & $\sim$19 & 5.1 & 99.94\% \\ %& 99.34\% \\
\hline
AP-3B &  0.80$\pm$0.31 & LSP & $\sim$12 & 5.4 & 99.76\% \\ %& 94.70\% \\
& & WWZ & $\sim$12 & 5.4 & 99.61\% \\ %& 99.27\%\\
\hline
AP-3C &  0.88$\pm$0.19 & LSP & $\sim$35 & 5.3 &  99.60\% \\ %& 88.78\% \\
& & WWZ & $\sim$34 & 5.4 & 99.54\% \\ %& 98.65\% \\
\hline
Q1 &  0.73$\pm$0.40 & LSP & $\sim$104 & 6.0 & 99.96\% \\ %& 96.37\% \\
& & WWZ & $\sim$104 & 6.0 & 99.93\% \\ %& 99.86\% \\
\hline 
Q2 &  0.60$\pm$0.26 & LSP & $\sim$227& 3.7 & 99.98\% \\ %& 99.33\% \\
& & WWZ & $\sim$223 & 3.7 & 99.96\% \\ %& 99.95\% \\
\hline\hline
\end{tabular}
\label{tab:1}
\end{table*}
The AP-3 phase lasted for 598 days (MJD 56150 - 56748) with an average flux of 0.37$\pm$0.01 unit. we found the hint of phases within it. We re-applied our phase identifying criteria as mentioned before to identify different sub-phases: AP-3A (MJD 56178 - 56274), AP-3B (MJD 56274 - 56400), AP-3C (MJD 56400 - 56591), and AP-3D (MJD 56591 - 56274) and then explored for recurrent signal in each. AP-3A and AP-3B show a periodicity of 19$\pm$2 days and 12$\pm$1.5 days, both with 5 complete cycles and a significance of 99.98\% and 99.76\% in LSP method (ref Table-\ref{tab:1}). AP-3C too shows a periodic flux variation from MJD 56406 to MJD 56591 with a periodicity of 35$\pm$4 days, but the result is at 99.60\% significance only in the LSP method.\\

\begin{figure*}
\centering
\includegraphics[width=0.45\textwidth,height=3.7cm]{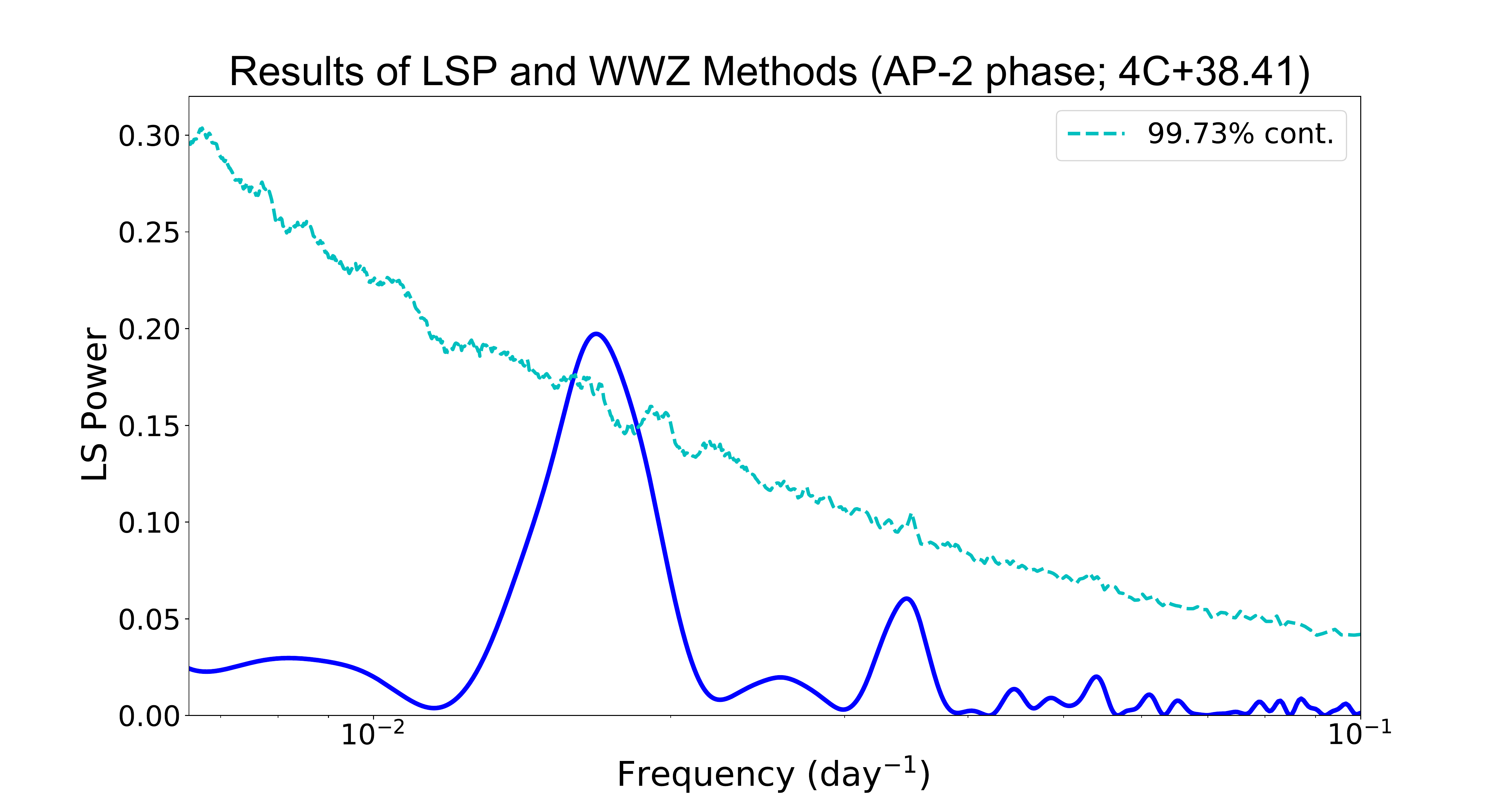}
\includegraphics[width=0.45\textwidth,height=3.7cm]{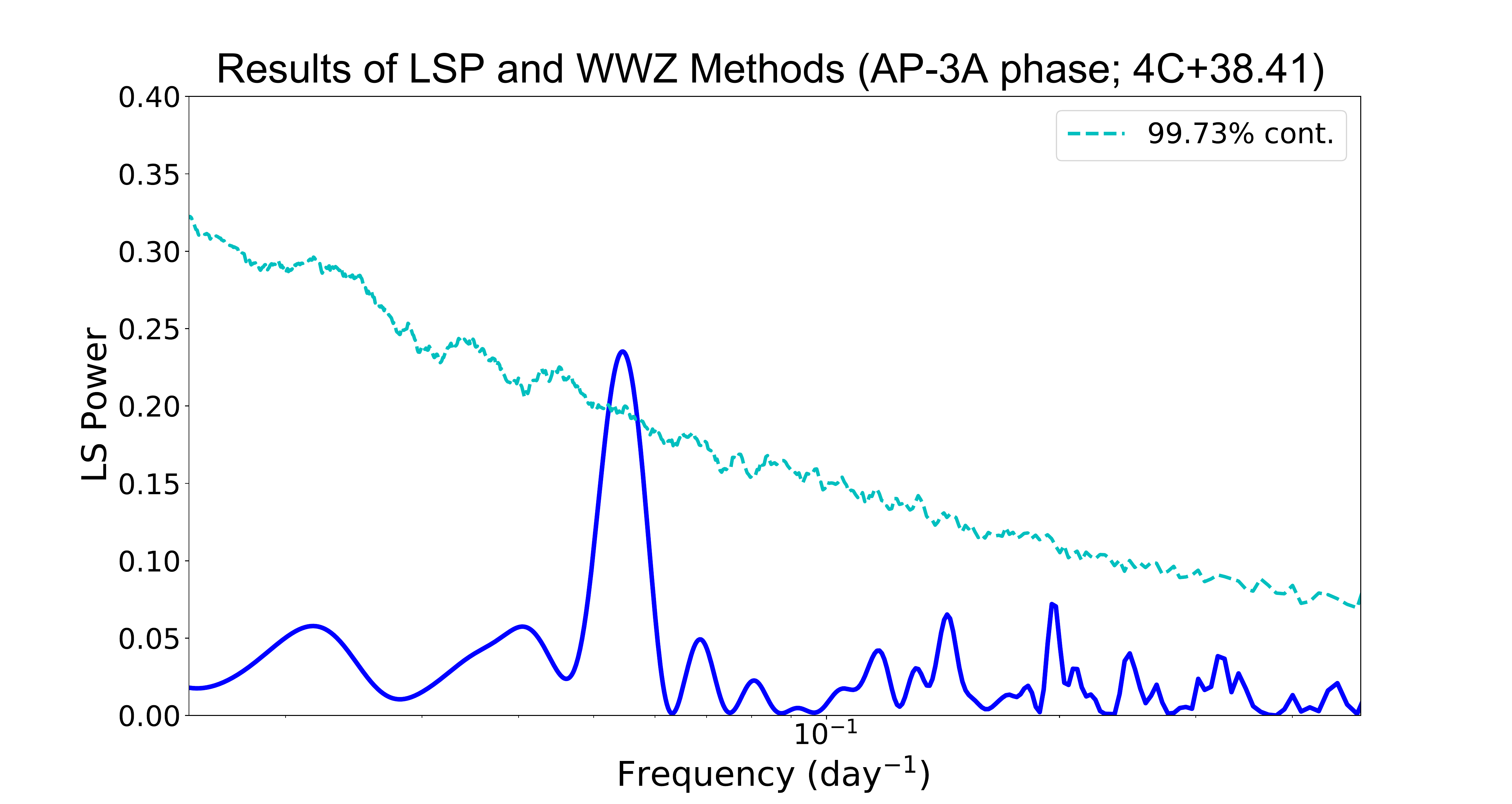}

\includegraphics[width=0.45\textwidth,height=3.7cm]{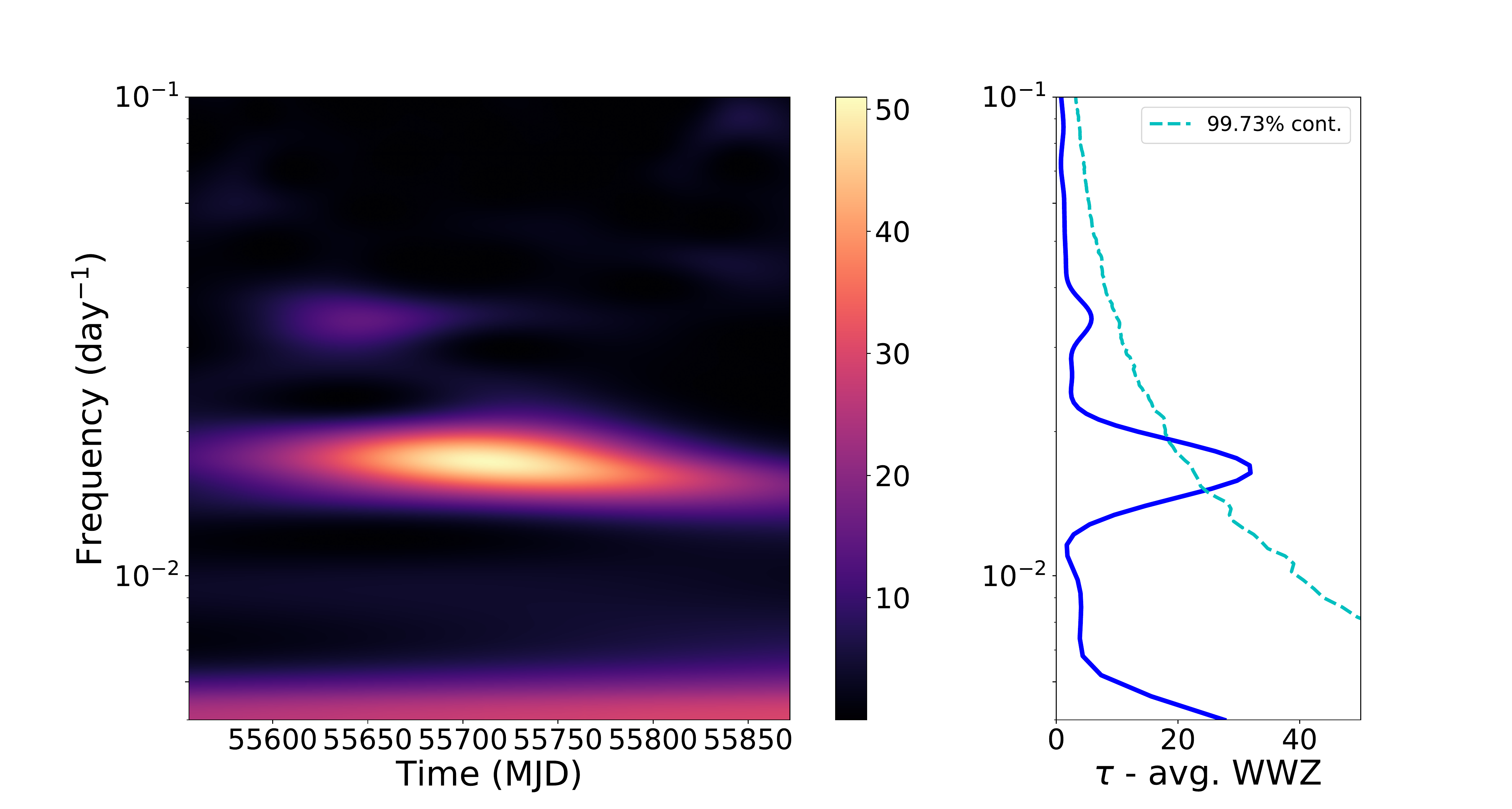}
\includegraphics[width=0.45\textwidth,height=3.7cm]{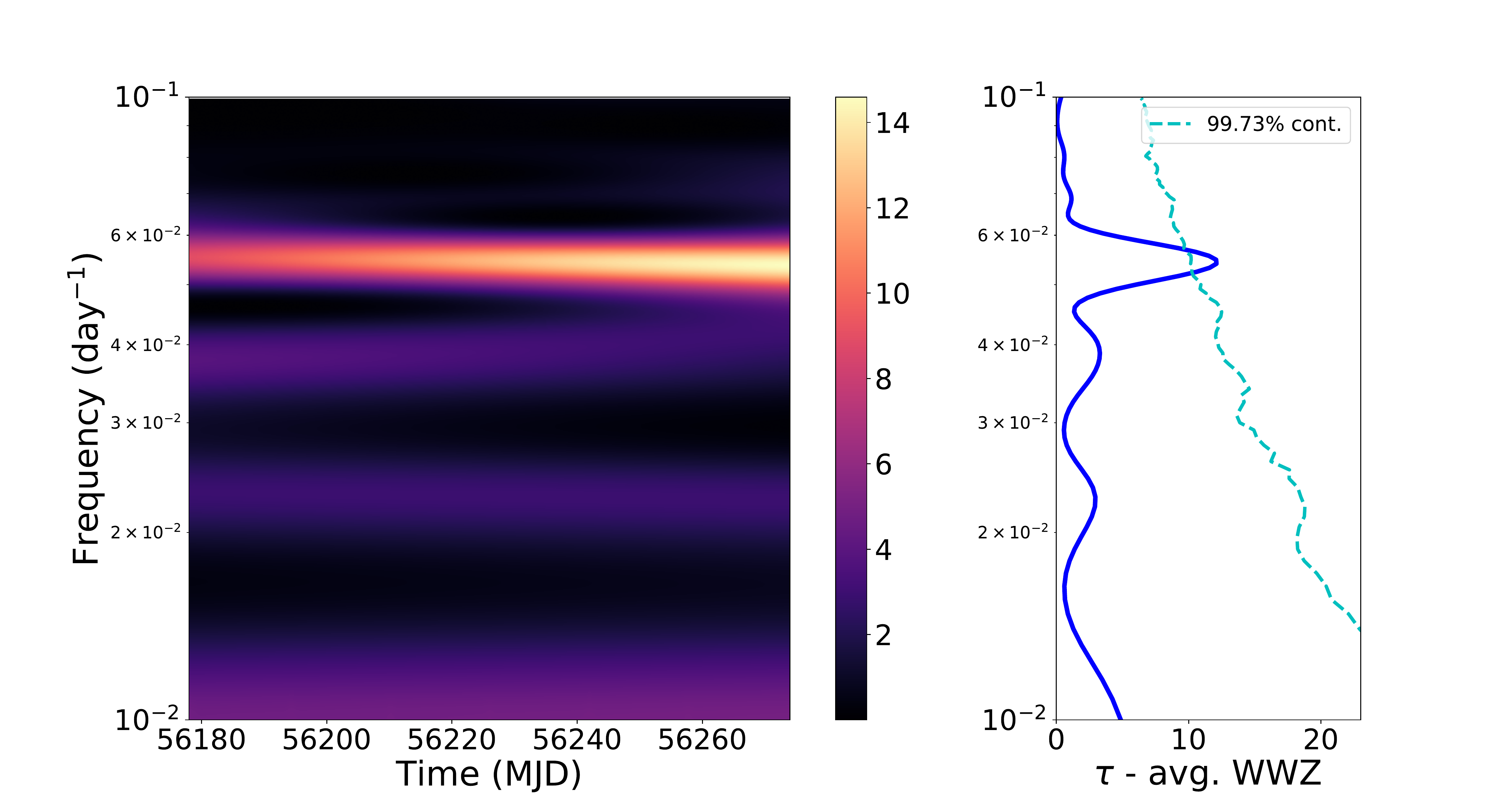}

\includegraphics[width=0.45\textwidth,height=3.7cm]{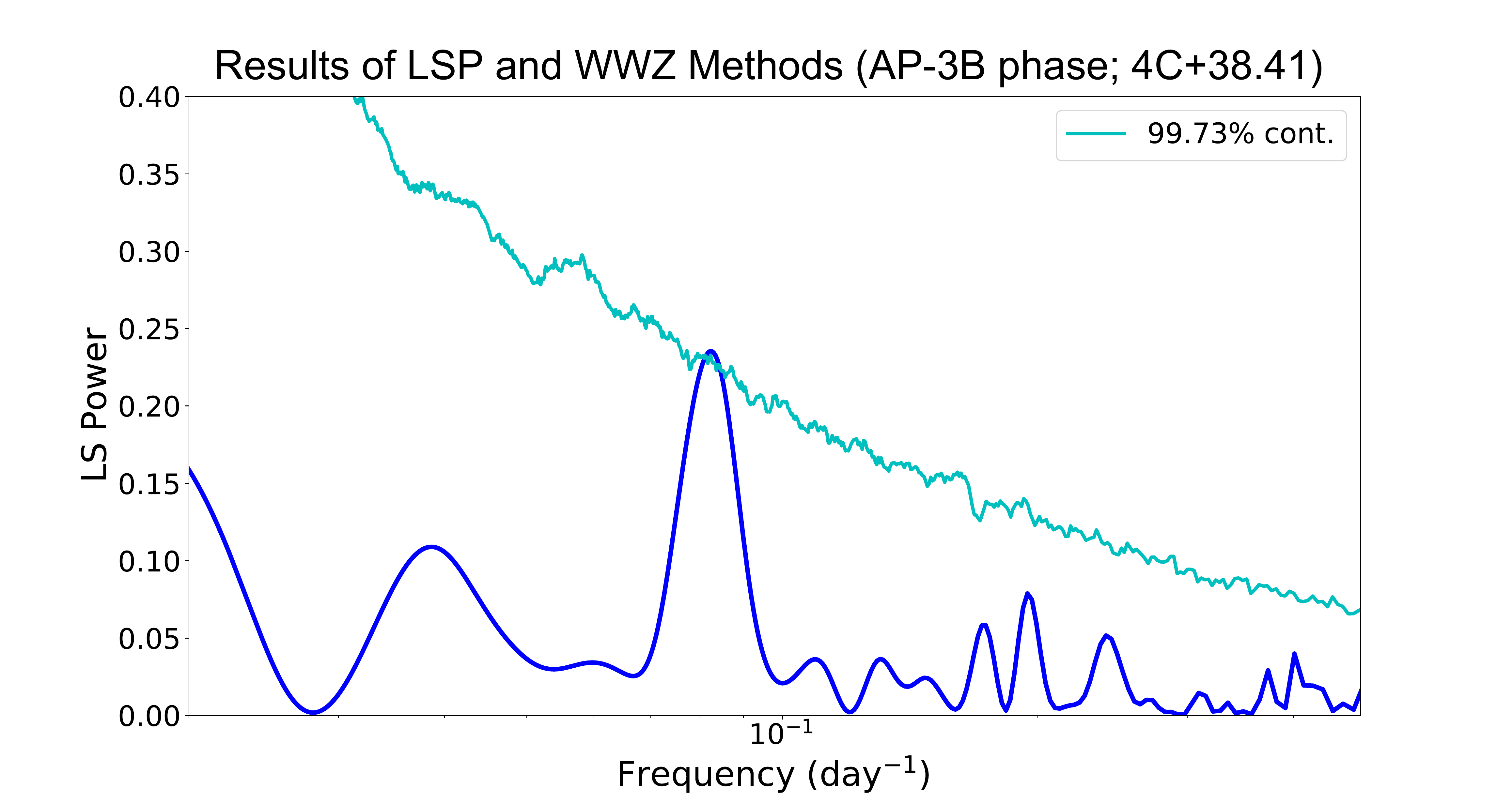}
\includegraphics[width=0.45\textwidth,height=3.7cm]{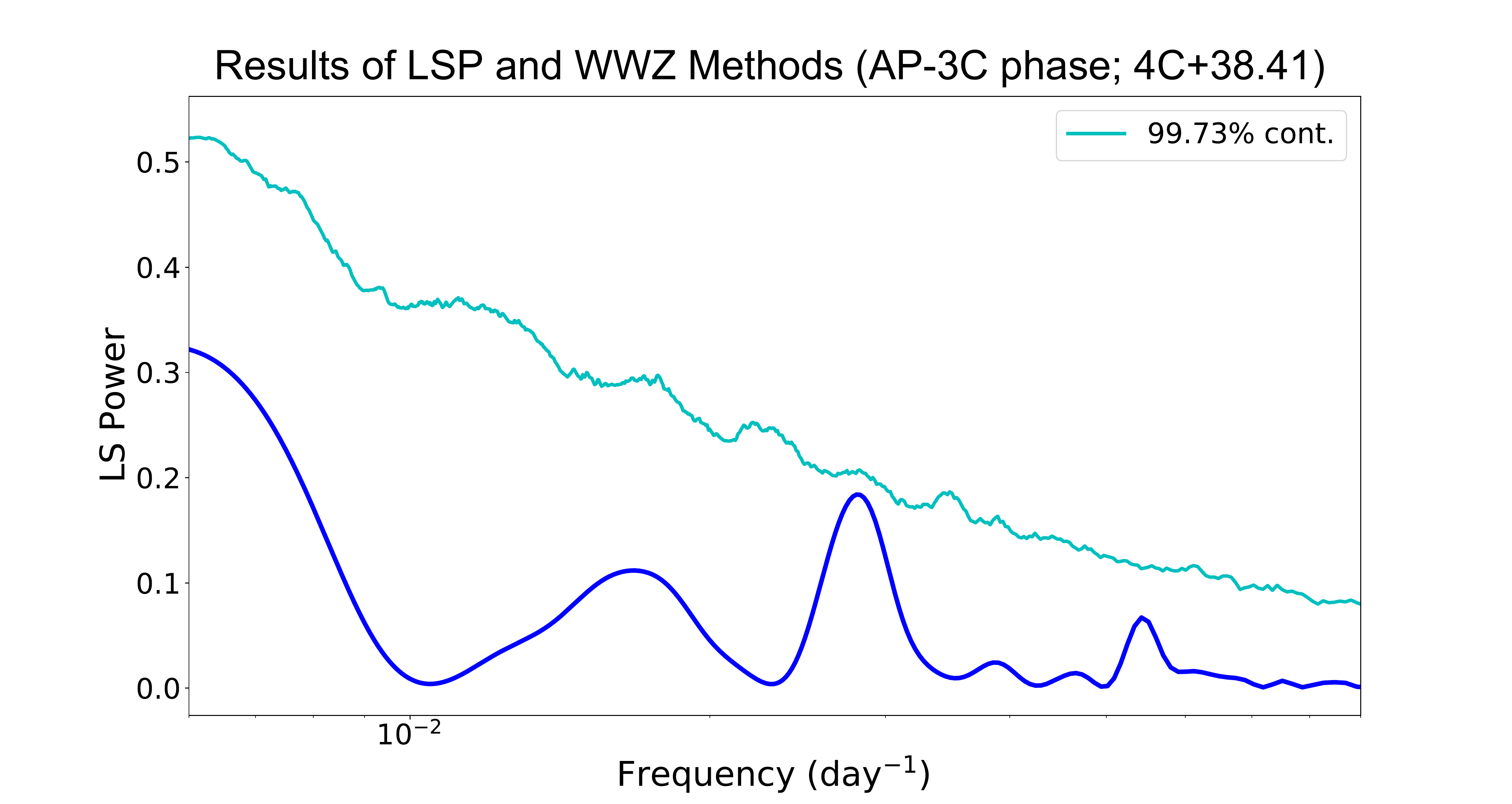}

\includegraphics[width=0.45\textwidth,height=3.7cm]{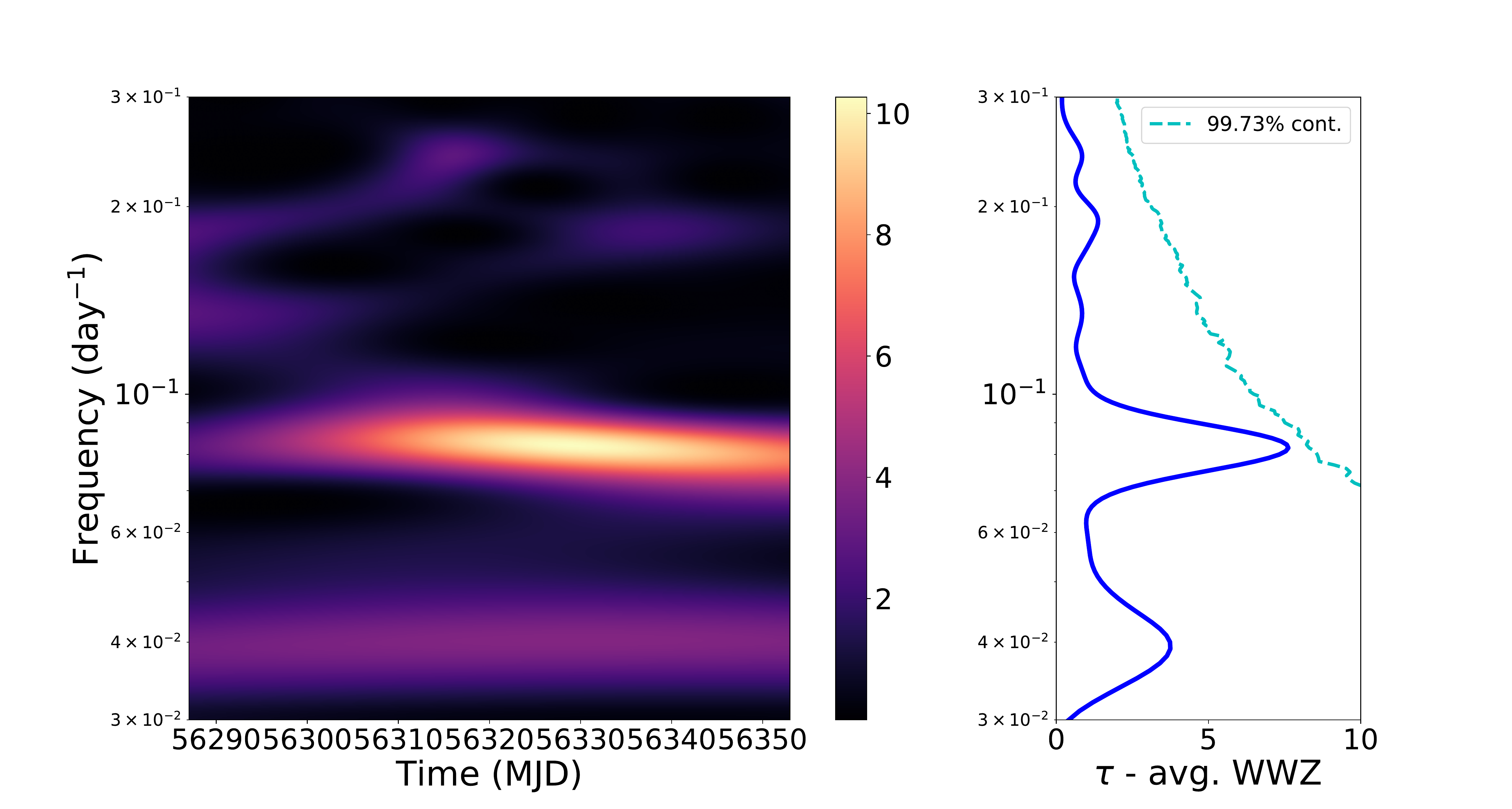}
\includegraphics[width=0.45\textwidth,height=3.7cm]{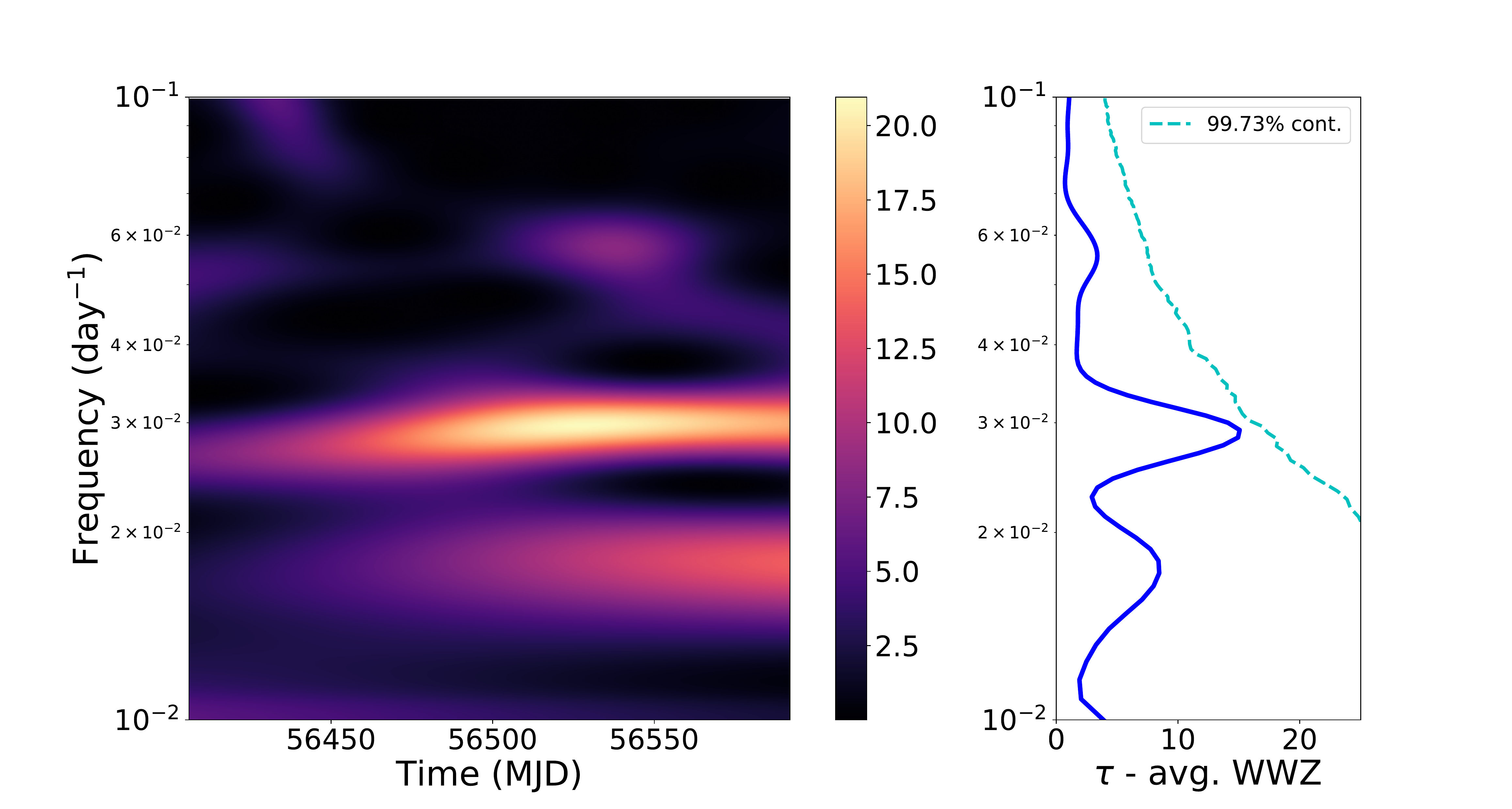}

\includegraphics[width=0.45\textwidth,height=3.7cm]{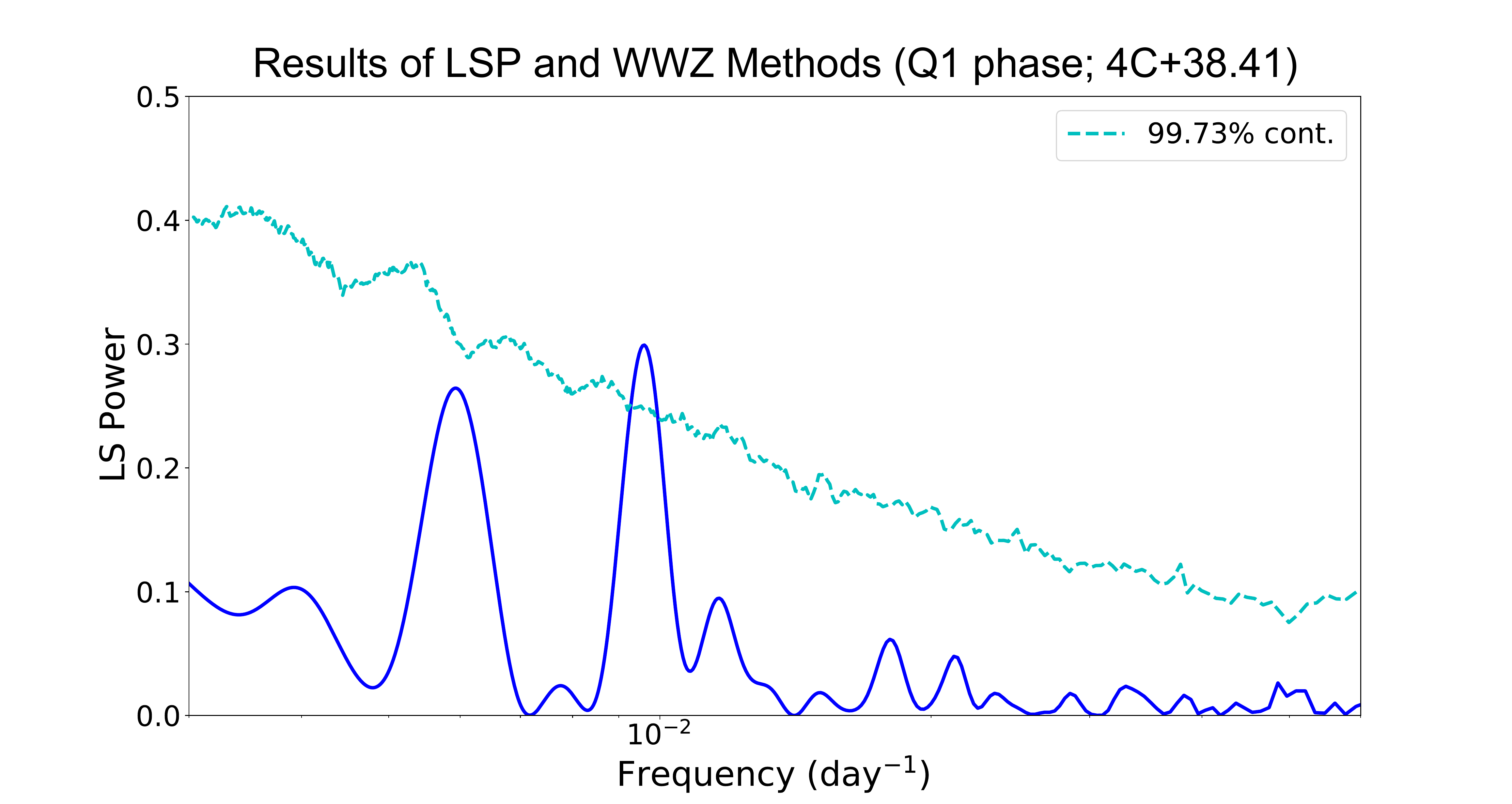}
\includegraphics[width=0.45\textwidth,height=3.7cm]{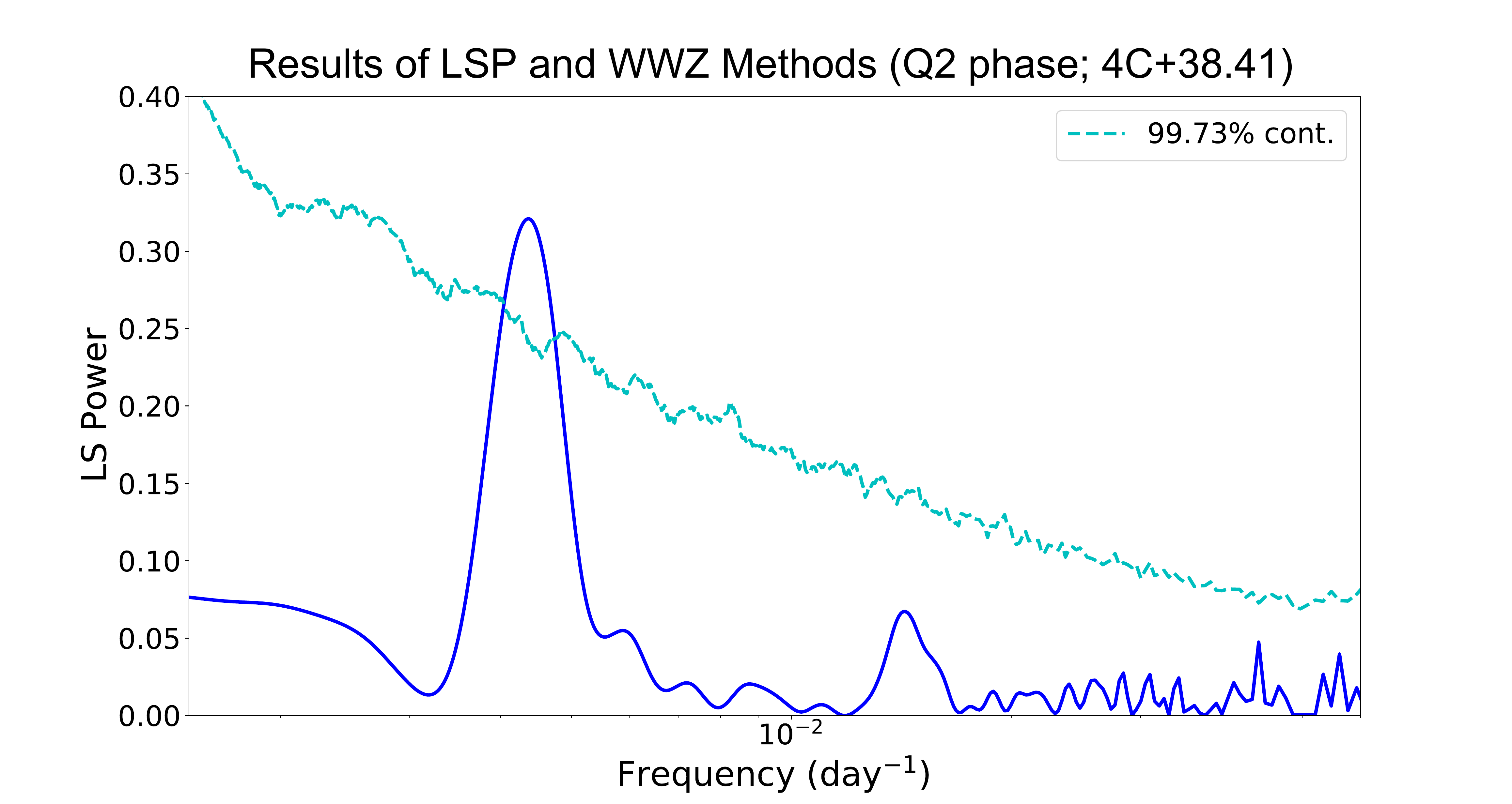}

\includegraphics[width=0.45\textwidth,height=3.7cm]{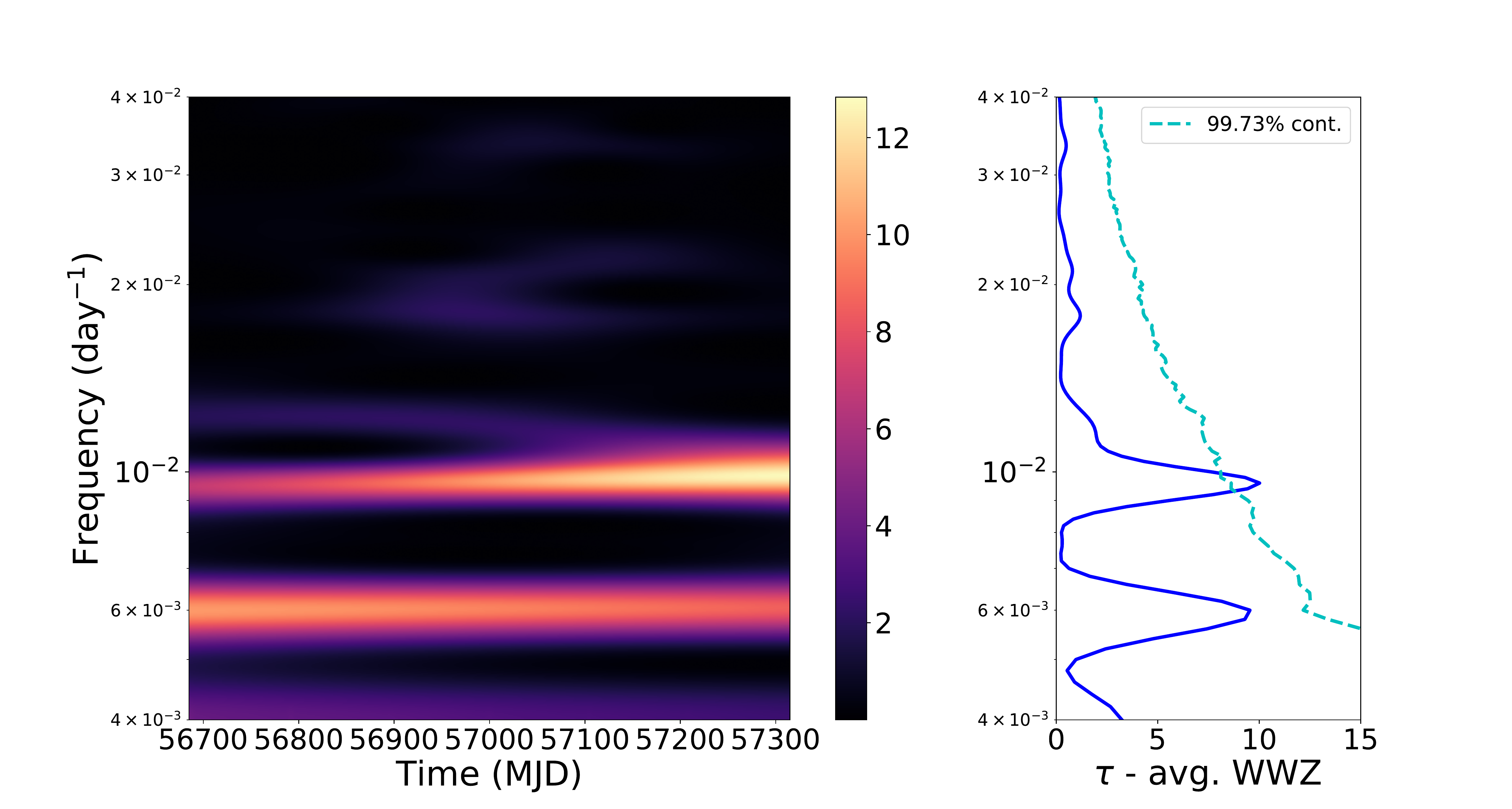}
\includegraphics[width=0.45\textwidth,height=3.7cm]{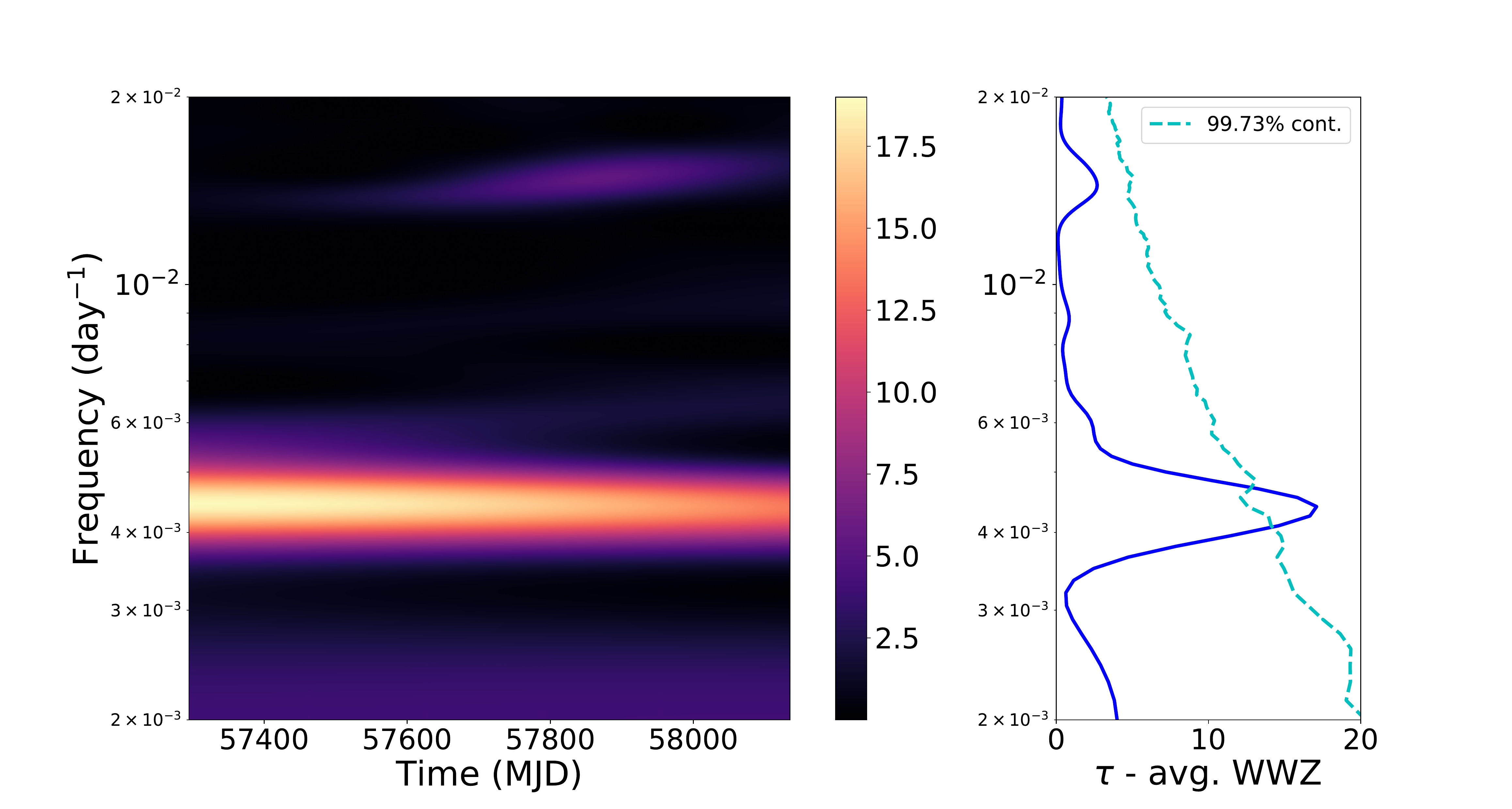}
\caption{Results of LSP and WWZ methods for different phases (AP-2, AP-3A, AP-3B, AP-3C, Q1, and Q2 phases of 4C+38.41). WWZ maps are shown below each phase of LSP plots (same as Figure-\ref{fig:5}). The cyan color curve represents the 99.73\% local significance contours in both methods.}
%\sout{along with the 99.73\% local significance curve in cyan color}
%95\% confidence contour is also shown in magenta color. 
\label{fig:6}
\end{figure*}

The quiescent phase (QP-1) of 4C+38.41 too has a few active episodes. The total time span of these phases is 1738 days (MJD 56598 - 58135) and 1070 days (MJD 58495 - 59565) for QP-1 and QP-2, respectively. QP-1 shows quasi-periodic-like behavior in two different time ranges. We used the same method to divide these time ranges as in the case of the AP-3 phase. We have defined this time range as Q1 and Q2 phases. Q1 phase (MJD 56685 - 57315) shows a periodicity of 104$\pm$7 days in 10 days binned light curve with greater than 99.93\% significance level in both the methods. Q2 phase (MJD 57305 - 58135) also shows QPO-like behavior (99.98\% in LSP and $\sim$ 99.96\% in WWZ method with a period of $\sim$ 227 days and $\sim$ 223 days respectively) but with nearly 3.7 cycles only. The QPO detection results (LSP and WWZ-maps) of all the phases of 4C+38.41 (except AP-1 phase) are shown in Figure-\ref{fig:6}. \\
%
%We have not found any QPO signature in Flare-3 (MJD 56150 - 56598). 
\\
AP-4 (MJD 58101 - 58497) shows QPO-like variation with a periodicity of 166 days. However, this result is relatively less significant (local significance level is 98.76\%), and the periodic nature lasts only for $\sim$ two cycles and thus, is not reliable from the stochastic nature point of view of time series. Due to this reason, we have not shown here the result of AP-4. On the other hand, given the source behavior during active episodes revealed by our study, this too could be a probable signal. The detailed outcomes of our study for all phases have also been given in Table-\ref{tab:1}\\
\\
We have also searched for QPOs in the entire 10-day binned light curve of 4C+38.41 and noticed three QPO-like features (957$\pm$117 days, 647$\pm$ 44 days, and 295$\pm$12 days), but none of them are significant (local significance level: $\sim$ 96.43\%, $\sim$ 99.07\%, $\sim$ 86.64\%), which is similar to the results presented by \citet{Bhatta2020} and \citet{Ren2022Apr}.\\

\section{Discussion} \label{sec:discuss}
We have explored temporal flux variability in $\sim$ thirteen years of Fermi-LAT data of the FSRQs PKS 0244-470 and 4C+38.41 in 10 days of binning (from MJD 54685 - 59565; reference Figure-\ref{fig:1}). These are both FSRQ-type blazars with $\gamma$-ray emission originating due to the inverse-Compton scattering of optical-UV photons in the leptonic scenario. The total gamma-ray emission can be the sum of the Comptionization of photon fields external to the jet including the accretion disk. If accretion-disk photon field contribution is significant (relative), then the observed signal could be associated with it. \\
\\
The QPOs study in blazars is one of the tools to know more about the central source, its surroundings, and the physical mechanisms responsible for the multi-wavelength emission. However, it requires well-sampled, good-quality data across the entire EM spectrum to clearly understand QPOs or their periodic nature. For instance, if the optical light curve also exhibits QPO-like variability correlated with the $\gamma$-ray emission, it suggests that the same relativistic electrons are responsible for both types of emissions. On the observational side, we do not see any QPO features in the optical band light curve taken from the Steward Observatory\footnote{\url{http://james.as.arizona.edu/ psmith/Fermi/}} (\citealt{Smith2009Dec}). Neither do we see any hint of QPO in the Steward polarimetric data  which is expected in kink-instability powered flares (\citealt{Dong2020May}). However, neither data is well-sampled and could be one of the reasons for the lack of QPOs in optical. Radio observations also play a crucial role in understanding the cause of significant QPOs or QPO-like variations in the jet. For PKS 0244-470, no radio data during the AP-1 phase are available and \citet{Algaba2018Jun} reported that 4C+38.41 is not resolvable in 129 GHz KVN (Korean VLBI Network) iMOGABA (Interferometric Monitoring of Gamma-ray Bright AGN) images due to its high redshift location. However, VLBA BU images \citep{Jorstad2017Sep} at 43 GHz (with a resolution of 0.2 mas) are available for this source (4C+38.41), and by using this data (from MJD 56000 – 57250), \citet{Algaba2018Jun} found that in only two cases, $\gamma$-ray flux enhancement occurred with the ejection of new radio components, which supports the shock-in-jet scenario \citep{Marscher1985Nov}. In other cases, no signature of new radio component ejection was found during $\gamma$-ray flares, suggesting a different origin of the flaring activity, e.g., variation of viewing angle and/or related to plasma instabilities \citep{Raiteri2012Sep}. So far, there is no clear evidence yet about the exact origin of $\gamma$-ray variability for this source and consequently, the observed transient QPO-like features. Therefore, further high-resolution VLBI observations, e.g., Event Horizon Telescope \citep{2019ApJ...875L...1E}, are needed in the future for a decisive interpretation. \\
\\
In our study, we have found transient QPOs or QPO-like variations in both sources with significance ranging from 99.60\% to 99.996\% level. It is important to note that the significance levels calculated here are strictly local, where we are interested in estimating the significance level for a particular frequency. However, without prior knowledge of the location of the peaks, it is more robust to check for a ``global significance."  In the ``global significance," we include the effect of observing our interested peak anywhere in the tested range (\citealt{Bell2011Feb,  Zhou2018Nov}). This effect is popularly known as the ``look-elsewhere effect" or ``multiple comparison problem" in statistics. We checked this for all the phases and found a reduction in the computed significance level. For example, the significance level of the AP-1 phase (PKS 0244-470) and Q1 phase (4C+38.41) are reduced to 98.164\% (99.996\% local significance) and 96.37\% (99.96\% local significance), respectively. These results are consistent with other recent works, e.g., \citealt{Covino2019Jan, Benkhali2020Feb}. However, in this regard, the QPO in PKS 0244-470 during AP-1 is different. If we neglect the last cycle (limiting to 7 cycles; from MJD 54685 - 56305) where it has weakened considerably, we get a very significant QPO with a local significance of 99.9997\% and a global significance of $\gtrsim$ 99.73\% (from simulating 3$\times$10$^{5}$ light curves) in the LSP method. The above discussions suggest that the claiming of the transient QPOs should be treated with caution. At the same time, it should be noted that very strict significance measures could fail to notice many exceptional/interesting features in AGN (see ``Result and Discussion" section in \citealt{Bhatta2020}), particularly in the low-frequency regime, and we may also miss exciting physics.\\
\\ 
%According to \citet{Algaba2018Jun}, 4C+38.41 is not resolvable in 129 GHz KVN (Korean VLBI Network) iMOGABA  (Interferometric Monitoring of Gamma-ray Bright AGN) images due to its high redshift location.
%its high redshift location and
\subsection{Physical Interpretations}
%\sout{In the past, the QPOs have been detected in many blazars and many possible explanations have been proposed to explain it depending upon the QPO time scale.} \sout{for these two sources}
QPOs in blazars can originate entirely within the jet or can be due to jet precession (\citealt{2004ApJ...615L...5R}) or binary SMBH scenario (\citealt{Ackermann2015}). However, binary SMBH and precession are expected to give long-term QPOs and thus are unable to provide satisfactory explanations as observed in our case. Below, we provide a few possible scenarios  which can explain the short-term periodicity (days to month-like) in the light curves.\\
\\
One of the well-known origins of the transient QPOs is the presence of a relativistic blob moving on a helical trajectory inside the jet. This blob can emit $\gamma$-ray radiation via External Compton (EC) and Synchrotron Self Compton (SSC) process (One-zone leptonic scenario). In this case, the time-dependent viewing angle ($\theta$) in the observer frame is given by \citep{Zhou2018Nov}:

\begin{equation} \label{eq:14}
    \cos{\theta(t)} = \cos{\phi} \cos{\psi} + \sin{\phi} \sin{\psi} \cos({2 \pi t/ P})
\end{equation}

Where $\phi$ and $\psi$ are the pitch angle (between blob's velocity direction and jet axis) and inclination angle (between jet axis and observer's line of sight), respectively. We have set $\phi$=2$^{\circ}$ and $\psi$ = 5$^{\circ}$ as in the case in \citet{Zhou2018Nov}. Using the typical value of Lorentz factor $\Gamma$ = 20 for FSRQs, we have computed the periodicity in the co-moving frame (P$^{'}$), distance traversed in one cycle of the helical motion (D$^{'}$), and total projected distance (S$^{'}$) for different phases of activity (Table-\ref{tab:2}). The main drawback of this model is that it can explain only the periodicity with constant amplitude. Several authors used a curved jet model to describe the varying amplitude scenario (\citealt{1992A&A...255...59C, 2021MNRAS.501...50S}), where inclination angle $\psi$ is time-dependent. However, the curved jet model finds it difficult to explain the short-term periodicity (e.g., AP-3A and AP-3B). We need high curvature for that, which is generally unusual for highly collimated, Mpc scale FSRQ jets. \\
%\sout{Value of different parameters from helical jet model (see text for more details). First column represents the different activity phases and quiescent phases of two FSRQs.}
\begin{table}
\centering
\caption{Helical jet parameters for mentioned features} %title of the table
%\resizebox{\columnwidth}{!}{\begin{tabular}{ccccc rrrrr} 
\begin{tabular}{ccccc rrrr}   % creating eight columns
\\
% & & Flare-A\\
\hline\hline                        
Activity & P & P$^{'}$ & D$^{'}$ & S$^{'}$ \\ [0.8ex] % inserts table 
%\hline\hline
& [days] & [years] & [pc] & [pc]\\
\hline\hline
& & PKS 0402-470 & & \\
\hline\hline
AP-1 & 225 & 109.0 & 33.4 & 23.3 \\
\hline\hline
& & 4C+38.41 & & \\
\hline\hline
AP-1 & 110 & 53.3 & 16 & 5.6 \\
\hline
AP-2 & 59 & 28.8 & 8.8 & 3.8 \\
\hline
AP-3A & 19 & 9.1 & 2.78 & 1.2 \\
\hline
AP-3B & 12 & 5.8 & 1.8 & 0.78 \\ 
\hline
AP-3C & 35 & 16.3 & 5.0 & 2.2 \\
\hline
Q1 & 104 & 49.9 & 15.3 & 8.0\\
\hline 
Q2 & 227 & 109.0 & 33.4 & 13.1\\
\hline\hline
% inserts single-line
\end{tabular}
%}
\label{tab:2}
\end{table}
\\
Another explanation of the QPO signature is given by \citealt{Dong2020May}. They have identified the blazar emission region inside the jet as the region of strongest kink instability. Due to these instabilities, there is a quasi-periodic conversion of magnetic energy to thermal energy. The observed period, in this case, can be given by:

\begin{equation} \label{eq:15}
    P = \frac{R_{KI}}{v_{tr}\delta}
\end{equation}

where $R_{KI}$ and $v_{tr}$ are the size of the emission region and transverse velocity respectively. $\delta$ is the Doppler factor of the jet. For typical blazar parameters value, they have found periodicities from week to month scale. They have also found in simulation that the polarization degree (PD) is anti-correlated with the variability of the light curve. However, for our sources, we cannot test this scenario due to the lack of good-quality PD data.\\
\\
QPO could also arise because of strong turbulent flow behind propagating shock \citep{1992vob..conf...85M}. The dominant turbulent cell exhibits enhanced Doppler boosting and can contribute the QPO component to the observed light curve at the turn-over period of the cell. However, due to the stochastic nature of the cell, it is highly likely that these QPOs last only for a few cycles \citep{Wiita2011Jun}. In our case, for AP-3A and AP-3B, this turn-over period is calculated as $\sim$ 68 days and $\sim$ 43 days respectively in the jet frame with the assumed Doppler factor, $\delta$ = 10.0 (\citealt{Raiteri2012Sep}, \citealt{Savolainen2010Mar}). This would require relatively reasonable-sized eddies to explain the observed QPOs \citep{Rani2009Apr}.\\
\\
Another plausible scenario is a bright hotspot revolving around the SMBH leading to enhanced production of $\gamma$-ray emission via the EC mechanism inside the jet. Due to Doppler boosting of the emission region, this model can explain the $\sim$ days like observed periodicity \citep{Roy2022Mar}. However, in blazars, the motion of this hotspot is symmetric around the jet axis, and thus it fails to explain the fast variability observed in the $\gamma$-ray light curve.\\
\\
Though we cannot identify the exact cause behind QPO-like variability as many processes seem consistent, kink instability seems the most probable candidate among all. Another indirect indication in favor of this is that almost all the strong activity phases have this QPO-like feature, indicating a similar underlying process. In the kink scenario, magnetic topology can give emission regions of varied sizes and the corresponding magnetic energy can be channeled in different proportions into bulk motion, particle injection, and/or acceleration giving rise to different observed variability time scales.  Even the relatively long-term (year-like) QPOs seen during several phases could result from this as current-driven kink instability is expected in magnetic-dominated regions, i.e., near the BH where the magnetic field is expected to be strong and weakened as one moves further from the BH. The occurrence of transient QPO-like variations, with timescales ranging from a few months to years, seems to be common in $\gamma$-ray bright blazars. For instance, a recent study by \citet{Ren2022Apr} claims that 24 out of 35 sources show such a trend. However, there is no study that examines short timescales (days to months) for this phenomenon. In contrast to other works, the occurrence of transient QPO-like variations in 4C+38.41 whenever the source enters a high state (except during AP-4) makes it an interesting source to study. This finding may imply that the kink instability keeps occurring in the jet with different sizes and/or velocities. However, it should be noted that these processes are highly complex and capable of producing a wide range of observational manifestations, and thus, more detailed and better data at the optical and radio band are required. For a longer period, a helical bend or helical jet model is also a possible explanation. However, this requires continuous particle acceleration/injection as the radiative cooling time in FSRQs is dominated by $\gamma$-ray and is of the order of a few minutes \citep[e.g.][]{2014MNRAS.442..131K}.

\section{Summary and Conclusions}
We explored the timing features in the $\sim$ 13 years long gamma-ray light curve of FSRQs 4C+38.41 and PKS 0244-470. We first identified the variability episodes and then employed the most widely used methods: LSP and WWZ to explore quasi-periodic features, followed by estimating their significance via the Monte Carlo approach using inputs from the observed light curve. The outcomes from our study is as follows 

\begin{itemize}
 \item Fermi-LAT analysis of the sources PKS 0244-470 and 4C+38.41 have been done with $\sim$ thirteen years (MJD 54800 - 59565) of archival data. Using Bayesian block with source intrinsic variability, we identified one activity and one quiescent phase in PKS 0244-470 and four activity and two Quiescent phases in 4C+38.41 with most phases showing recurring features.\\ 
 \item AP-1 phase of PKS 0244-470 shows QPO-like behavior with a period of $\sim$ 225 days and persists for nearly eight cycles. This is the first time we have reported this significant result: 99.996\% and 99.986\% local significance level in LSP and WWZ methods, respectively. This feature remains significant globally too. \\
 \item All excpet one (AP-4) of the activity phases of 4C+38.41 show QPO-like behavior in their light curve with four to five complete cycles. The duration range from $\sim 10 - 110$ days. One of the phases shows three sub-phases, each showing a QPO-like feature, ranging from $\sim$ 10 to 40 days. \\
 \item  The Quiescent phases of 4C+38.41 also show QPO-like behavior on two different times scales: $\sim$ 104 days (six cycles) and $\sim$ 223 days (nearly 3.7 cycles) with local significance level $\gtrsim$ 99.93\%. \\
 %\sout{Q1 phase (MJD 56685 - 57315) shows a periodicity of $\sim$ 104 days with six complete cycles. On the other hand Q2 phase (MJD 57295 - 58315) has a periodicity of $\sim$ 223 days (nearly 3.7 cycles) in the light curve. The local significance level of both results is above 99.93\%.}
 
 \item Kink-instability seems the most probable explanation for both short and long-term observed QPOs. However, the curved jet model can also explain the relatively longer period QPOs. \\
 
\end{itemize}

The global significance of these transient QPO-likes features is within the range reported by researchers in other blazars.

\section*{Acknowledgements}

We thank the anonymous referee for his/her comments and suggestions which have helped us to improve the manuscript. A.K. Das thanks Gopal Bhatta for fruitful discussions and S.K. Mondal for providing the Fermi-LAT data for one observation set. R.P. acknowledges the support of the Polish Funding Agency National Science Centre, project 2017/26/A/ST9/-00756 (MAESTRO 9) and the European Research Council (ERC) under the European Union’s Horizon 2020 research and innovation program (grant agreement No. [951549]). P.K. acknowledges support from the Department of Science and Technology (DST), Government of India, through the DST-INSPIRE faculty grant (DST/
INSPIRE/04/2020/002586).
\\
%%%%%%%%%%%%%%%%%%%%%%%%%%%%%%%%%%%%%%%%%%%%%%%%%%
\section*{Data Availability}

This work has made use of publicly available Fermi-LAT data obtained from FSSC's website data server and provided by NASA Goddard Space Flight Center. Photon index data is taken from Fermi LAT Light Curve Repository webpage \citep{2021ATel15110....1F}.
\\
\software{ 
Fermitools (\href{https://fermi.gsfc.nasa.gov/ssc/data/analysis/scitools/}{https://fermi.gsfc.nasa.gov/ssc\\/data/analysis}), Fermi-LAT Light Curve Respiratory (\href{https://fermi.gsfc.nasa.gov/ssc/data/access/lat/LightCurveRepository}{https://fermi.gsfc.nasa.gov/lat/LightCurveRepository}), WWZ method (\href{https://github.com/eaydin/WWZ/}{https://github.com/eaydin/WWZ}) } 

\bibliographystyle{plain}

\bibliography{QPO_4C+38}{}
\bibliographystyle{aasjournal} 
%\appendix
%%%%%%%%%%%%%%%%%%%%%%%%%%%%%%%%%%%%%%%%%%%%%%%%%%

\end{document}